\documentclass[a4paper,12pt]{article}
\usepackage{amsmath,enumerate}
\usepackage{graphicx}

\newcommand{\lapprox}{%
\mathrel{%
\setbox0=\hbox{$<$}
\raise0.6ex\copy0\kern-\wd0
\lower0.65ex\hbox{$\sim$}
}}
\newcommand{\gapprox}{%
\mathrel{%
\setbox0=\hbox{$>$}
\raise0.6ex\copy0\kern-\wd0
\lower0.65ex\hbox{$\sim$}
}}

\begin{document}

\begin{center}

{\Large \bf Lepton flavor violating $Z$ and Higgs decays in the scotogenic
model}\\[20mm]

Raghavendra Srikanth Hundi\\
Department of Physics, Indian Institute of Technology Hyderabad,\\
Kandi - 502 284, India.\\[5mm]
E-mail address: rshundi@phy.iith.ac.in \\[20mm]

\end{center}

\begin{abstract}

In this work, we have studied lepton flavor violating (LFV) decays of $Z$
gauge boson
and Higgs boson ($H$) in the scotogenic model. We have computed branching
ratios for the decays $Z\to\ell_\alpha\ell_\beta$ and
$H\to\ell_\alpha\ell_\beta$ in this model. Here, $\ell_\alpha$ and
$\ell_\beta$ are different charged lepton fields. After fitting to
the neutrino oscillation observables in the scotogenic model, we have found
that the branching
ratios for the LFV decays of $Z$ and $H$ can be as large as $\sim10^{-8}$ and
$\sim10^{-3}$ respectively. However, after satisfying the constraints due to
non-observation of $\ell_\alpha\to\ell_\beta\gamma$ decays, the above mentioned
branching ratio results are found to be suppressed by a factor of
$\sim10^{-7}$.

\end{abstract}
\newpage

\section{Introduction}

Physics beyond the standard model \cite{bsm} can be probed by searching
for lepton
flavor violating (LFV) \cite{lfvrev} processes in experiments. So far no
LFV signal has
been observed in experiments, and as result, upper bounds exist on various
LFV processes \cite{pdg}. In the standard model these experimental limits
are satisfied, since LFV
processes are highly suppressed due to Glashow-Iliopoulos-Maiani cancellation
mechanism. On the other hand, in a beyond standard model, the branching
ratios for these processes
can be appreciably large and the model can be constrained by experiments.

Scotogenic model \cite{scoto} is an extension of standard model, which
explains the neutrino mass and dark matter problems, which are briefly
described below. Neutrino masses are
found to be tiny \cite{osci}, and hence, in order to explain the smallness
of neutrino masses a different mechanism should be proposed for them
\cite{nurev}. Regarding the dark matter problem, it is known that the
universe consists of nearly 25$\%$ of energy in the form of non-baryonic
matter \cite{dmrev}, which cannot be explained by the standard model.
In the scotogenic model, the origin of neutrino masses are explained by a
radiative mechanism by proposing an extra scalar doublet ($\eta$),
three right-handed Majorana neutrinos ($N_k$) and an additional $Z_2$ symmetry.
Under $Z_2$ symmetry, which is unbroken, $\eta$ and $N_k$ are odd and all
the standard model
fields are even. As a result of this, the lightest among the neutral
$Z_2$-odd particles can be a candidate for the dark matter.

Various phenomenological consequences of scotogenic model have been studied
in relation to LFV, dark matter, matter-antimatter asymmetry and
colliders \cite{kms,tv,pheno}. In the studies on LFV in the scotogenic model,
the following processes have been analyzed:
$\ell_\alpha\to\ell_\beta\gamma$,
$\ell_\alpha\to3\ell_\beta$ and conversion of $\mu$ to $e$
\cite{kms,tv}.
In a related direction, see Ref. \cite{prwo}, for a study on LFV in the
supersymmetric scotogenic model \cite{susyscoto}. In contrast to above
mentioned studies, in this work, we analyze
the LFV decays of $Z$ and Higgs boson in the scotogenic model \cite{scoto}.
The decays $Z\to\ell_\alpha\ell_\beta$ and $H\to\ell_\alpha\ell_\beta$ are
driven at
1-loop level by $\eta^\pm$ and $N_k$, where $\eta^\pm$ is the charged component
of $\eta$. We compute branching ratios for these decays, which we find
to be dependent on
the Yukawa couplings and masses of $\eta^\pm$ and $N_k$. By varying the
parameters of the model, we study on the reach of the values of the above
mentioned branching ratios.

The current experimental
bounds on the branching ratios of $Z\to\ell_\alpha\ell_\beta$ and
$H\to\ell_\alpha\ell_\beta$ are as follows.
\begin{eqnarray}
{\rm Br}(Z\to e\mu)&<&7.5\times10^{-7}~\cite{zemu},
\nonumber \\
{\rm Br}(Z\to e\tau)&<&9.8\times10^{-6}~\cite{zetau},
\nonumber \\
{\rm Br}(Z\to \mu\tau)&<&1.2\times10^{-5}~\cite{zmutau}.
\label{zexp}
\end{eqnarray}
\begin{eqnarray}
{\rm Br}(H\to e\mu)&<&6.1\times10^{-5}~\cite{hemu},
\nonumber \\
{\rm Br}(H\to e\tau)&<&4.7\times10^{-3}~\cite{hetau},
\nonumber \\
{\rm Br}(H\to \mu\tau)&<&2.5\times10^{-3}~\cite{hmutau}.
\label{hexp}
\end{eqnarray}
In future, LFV decays of $Z$ and $H$ will be probed. For instance, in the
upcoming $e^+e^-$ collider such as the FCC-ee, the following sensitivites
can be probed for the LFV decays of $Z$ \cite{tesla}.
\begin{eqnarray}
{\rm Br}(Z\to e\mu)&\sim&10^{-10}-10^{-8},
\nonumber \\
{\rm Br}(Z\to e\tau)&\sim&10^{-9},
\nonumber \\
{\rm Br}(Z\to \mu\tau)&\sim&10^{-9}.
\label{fzexp}
\end{eqnarray}
Similarly, the bounds on LFV decays of Higgs
boson, given in Eq. (\ref{hexp}), may be reduced in future by the LHC. Since
in future experiments, there is an interest to probe LFV decays of $Z$ and $H$,
it is worth to compute the branching ratios of these decays in the scotogenic
model. It is also interesting to analyze the status of the above mentioned
decays in this model, in relation to the present and future bounds on them.

As already stated, the LFV decays of $Z$ and $H$ are mediated at 1-loop
by $\eta^\pm,N_k$ in the scotogenic model. The same mediating particles,
in this model, can also drive $\ell_\alpha\to\ell_\beta\gamma$ at 1-loop level.
As a result of this, there exist a correlation between branching ratios
of $Z,H\to\ell_\alpha\ell_\beta$ and that of $\ell_\alpha\to\ell_\beta\gamma$.
Since stringent bounds exists on the non-observation of
$\ell_\alpha\to\ell_\beta\gamma$ \cite{pdg}, we
have studied the implications of those bounds on the branching ratios of
$Z,H\to\ell_\alpha\ell_\beta$. For related studies on LFV decays of $Z$ and $H$,
see Refs.\cite{zdec,hdec}.

The neutrino masses in the scotogenic model are generated at 1-loop level
through the mediation of neutral components of $\eta$ and $N_k$. As a result
of this, neutrino masses in this model depend on neutrino Yukawa
couplings and masses of neutral components of $\eta$ and $N_k$.
As already stated before, the
branching ratios for $Z,H\to\ell_\alpha\ell_\beta$ also depend on the
neutrino Yukawa couplings and masses of $\eta^\pm$ and $N_k$. One can notice
that there exist a correlation between branching ratios of
$Z,H\to\ell_\alpha\ell_\beta$ and neutrino masses and mixing angles.
We have explored this correlation and we have found that the branching ratios
of $Z\to\ell_\alpha\ell_\beta$ can reach as high as $10^{-8}$ by satisfying
the perturbativity limits on the parameters of the scotogenic model. On
the other hand, the branching ratios for $H\to\ell_\alpha\ell_\beta$ can
reach as high as $10^{-3}$. However, the above mentioned results are obtained
without imposing the constraints due to non-observation of
$\ell_\alpha\to\ell_\beta\gamma$.
After imposing the constraints due to $\ell_\alpha\to\ell_\beta\gamma$, we
have found that the above mentioned results on the branching ratios are
suppressed by a factor of $10^{-7}$. As a result of this, the decay
$H\to\mu\tau$ is found to
have the highest branching ratio of $\sim10^{-10}$, in our analysis on the
LFV decays of $Z$ and $H$.

In this work, although we study LFV decays of both $Z$ and $H$,
only the LFV decays of $H$ have been studied in
Ref. \cite{hrs}. Our method of
computing the branching ratios for LFV decays of $H$ is different from that
of Ref. \cite{hrs}. Moreover, only an estimation on the branching
ratio of $H\to\mu\tau$ has been made in Ref. \cite{hrs},
in the context of scotogenic model. Whereas, we
have studied branching ratios for all LFV Higgs decays in more details here.
We compare our results with
that of Ref. \cite{hrs} towards the end of this paper. See
Ref. \cite{gsc} for some discussion on LFV decays of $Z$ and $H$ in
the context of generalized scotogenic model.

The paper is organized as follows. In the next section, we briefly describe
the scotogenic model. In Sec. 3, we present analytic expressions on the
branching ratios of $Z\to\ell_\alpha\ell_\beta$ and
$H\to\ell_\alpha\ell_\beta$ in the scotogenic model. In Sec. 4, we
analyze these branching ratios and present our numerical results on them.
We conclude in the last section.

\section{Scotogenic model}

Scotogenic model \cite{scoto} is an extension of the standard model, where the
additional fields are one $SU(2)$ scalar doublet $\eta=(\eta^+,\eta^0)^T$ and
three singlet right-handed
neutrinos $N_k$. This model has an additional discrete $Z_2$ symmetry,
under which $\eta,N_k$ are odd and all the standard model fields are even.
To construct the invariant Lagrangian of this model, we can choose a basis
where the Yukawa couplings for charged leptons and the masses of right-handed
neutrinos are diagonal. In such a basis, the Lagrangian of this model in the
lepton sector is \cite{scoto}
\begin{equation}
-{\cal L}_Y=f_\alpha\bar{L}_{L\alpha}\phi \ell_{R\alpha}+h_{\alpha k}
\bar{L}_{L\alpha}\eta^cN_k+\frac{M_k}{2}\overline{N^c_k}N_k+h.c.
\label{lag}
\end{equation}
Here, $\alpha=e,\mu,\tau$ and $k=1,2,3$.
$L_{L\alpha}=(\nu_{L\alpha},\ell_{L\alpha})^T$ is a left-handed lepton
doublet, $\ell_{R\alpha}$ is a right-handed singlet charged lepton,
$\phi=(\phi^+,\phi^0)^T$ is the scalar Higgs doublet and
$\eta^c=i\sigma_2\eta^*$, where $\sigma_2$ is a Pauli matrix.
$\phi$ and $\eta$ are the only two scalar fields of this model. The scalar
potential between these two fields is given below \cite{scoto}.
\begin{eqnarray}
V&=&m_1^2\phi^\dagger\phi+m_2^2\eta^\dagger\eta+\frac{1}{2}\lambda_1
(\phi^\dagger\phi)^2+\frac{1}{2}\lambda_2(\eta^\dagger\eta)^2
+\lambda_3(\phi^\dagger\phi)(\eta^\dagger\eta)+\lambda_4(\phi^\dagger\eta)
(\eta^\dagger\phi)
\nonumber \\
&&+\frac{1}{2}\lambda_5[(\phi^\dagger\eta)^2+h.c.].
\end{eqnarray}
Here, $\lambda_5$ is chosen to be real, without loss of generality. Since $Z_2$
is an exact symmetry of this model, we should have $m_1^2<0$ and $m_2^2>0$
so that only $\phi$ acquires vacuum expectation value (VEV), whereas $\eta$
does not acquire VEV.
Since only $\phi$ acquires VEV, the physical fields in the neutral components
of $\phi$ and $\eta$ can be written as
\begin{equation}
\phi^0=\frac{H}{\sqrt{2}}+v,\quad\eta^0=\frac{1}{\sqrt{2}}(\eta_R+i\eta_I)
\end{equation}
Here, $H$ is the Higgs boson and $v\approx$ 174 GeV. Now, after the electroweak
symmetry breaking, the physical components of $\phi$ and $\eta$ acquire
masses, whose expressions in the form of mass-squares are given below
\cite{scoto}.
\begin{eqnarray}
m^2(H)\equiv m_H^2&=&2\lambda_1v^2,
\nonumber \\
m^2(\eta^\pm)\equiv m_{\eta^\pm}^2&=&m_2^2+\lambda_3v^2,
\nonumber \\
m^2(\eta_R)\equiv m_R^2&=&m_2^2+(\lambda_3+\lambda_4+\lambda_5)v^2
=m_0^2+\lambda_5v^2,
\nonumber \\
m^2(\eta_I)\equiv m_I^2&=&m_2^2+(\lambda_3+\lambda_4-\lambda_5)v^2
=m_0^2-\lambda_5v^2
\end{eqnarray}
Here, $m_0^2=m_2^2+(\lambda_3+\lambda_4)v^2$.

After the electroweak symmetry breaking, the first term of Eq. (\ref{lag})
give masses to charged leptons, whose expressions can be written as
\begin{equation}
m_{\ell_\alpha}=f_\alpha v
\end{equation}
On the other hand, since $\eta$ does not acquire VEV, the second term of
Eq. (\ref{lag}) do not generate Dirac masses for neutrinos. As a result
of this, neutrinos are massless at tree level. However, at 1-loop level,
neutrinos acquire masses through the mediation of neutral components of
$\eta$ and $N_k$ \cite{scoto}. By taking $\Lambda={\rm diag}(\Lambda_1,
\Lambda_2,\Lambda_3)$, the mass expressions for neutrinos at 1-loop level
can be written as follows \cite{scoto}.
\begin{eqnarray}
(M_\nu)_{\alpha\beta}&=&(h\Lambda h^T)_{\alpha\beta}=\sum_{k=1}^3
h_{\alpha k}h_{\beta k}\Lambda_k,
\nonumber \\
\Lambda_k&=&\frac{M_k}{16\pi^2}\left[\frac{m_R^2}{m_R^2-M_k^2}
\ln\frac{m_R^2}{M_k^2}-\frac{m_I^2}{m_I^2-M_k^2}\ln\frac{m_I^2}{M_k^2}\right]
\label{numas}
\end{eqnarray}
Using the Casas-Ibarra parametrization \cite{ci}, the matrix
containing Yukawa couplings $h_{\alpha k}$ can be parametrized as
\begin{equation}
h=U_{PMNS}^*\sqrt{m_\nu}R\sqrt{\Lambda}^{-1}
\label{cipara}
\end{equation}
Here, $U_{PMNS}$ is the Pontecorvo-Maki-Nakagawa-Sakata matrix, which can be
parametrized \cite{pdg} in terms of the three neutrino mixing angles,
one $CP$ violating Dirac phase and two Majorana phases.
$m_\nu$ is a diagonal matrix containing the neutrino mass eigenvalues,
which can be written as $m_\nu={\rm diag}(m_1,m_2,m_3)$. $R$ is a complex
orthogonal matrix which satisfies $RR^T=I=R^TR$. Using the parametrization
of Eq. (\ref{cipara}), one can notice that
\begin{equation}
M_\nu=U_{PMNS}^*m_\nu U_{PMNS}^\dagger
\end{equation}
From the above equation, we can see that the unitary matrix which diagonalize
$M_\nu$ is $U_{PMNS}$. Hence, the mixing pattern in the neutrino sector of
the scotogenic model can be explained by parametrizing the Yukawa couplings
as given by Eq. (\ref{cipara}).

As described in Sec. 1, the aim of this work is to analyze LFV decays of
$Z$ and $H$. One can notice that the LFV processes in the scotogenic
model are driven by the off-diagonal Yukawa couplings of the second term of
Eq. (\ref{lag}). In the next section, we explicitly show that the branching
ratios of the LFV decays for $Z$ and $H$ are proportional to off-diagonal
elements of $h_{\alpha k}$. As a result of this, the above mentioned branching
ratios are unsuppressed if $h_{\alpha k}\sim1$. On the other hand,
$h_{\alpha k}$ also determine neutrino masses from Eq. (\ref{numas}). As
already pointed in Sec. 1, masses of neutrinos are very small. Hence,
in order to explain the smallness of neutrino masses along
with $h_{\alpha k}\sim1$, one can make $\Lambda_k$ very small.
The above statement is possible if one takes $m_R^2$ and $m_I^2$ to be
nearly degenerate, which is described below. In this work, we take the
masses of the components of
$\eta$ and $M_k$ to be around few hundred GeV. Now, after using
$\lambda_5\ll1$ in the expressions for $m_R^2$ and $m_I^2$, up to
first order in $\lambda_5$, we get
\begin{equation}
\Lambda_k=\frac{M_k}{8\pi^2}\frac{\lambda_5v^2}{m_0^2-M_k^2}\left[1-
\frac{M_k^2}{m_0^2-M_k^2}\ln\frac{m_0^2}{M_k^2}\right]
\end{equation}
Using the above equation, one can notice that the smallness of neutrino
masses in the scotogenic model can be explained by suppressing the
$\lambda_5$ coupling. For this choice of $\lambda_5$, the Yukawa couplings
$h_{\alpha k}$ are ${\cal O}(1)$, which can lead to unsuppressed decay
rates for LFV processes in the scotogenic model.

\section{Analytic expressions for the branching ratios of
$Z\to\ell_\alpha\ell_\beta$ and $H\to\ell_\alpha\ell_\beta$}

In the scotogenic model, the LFV decays of $Z$ and $H$ are dominantly
driven by $\eta^\pm$ and $N_k$, which are shown in Fig. 1.
\begin{figure}[h]
\begin{center}

\includegraphics[]{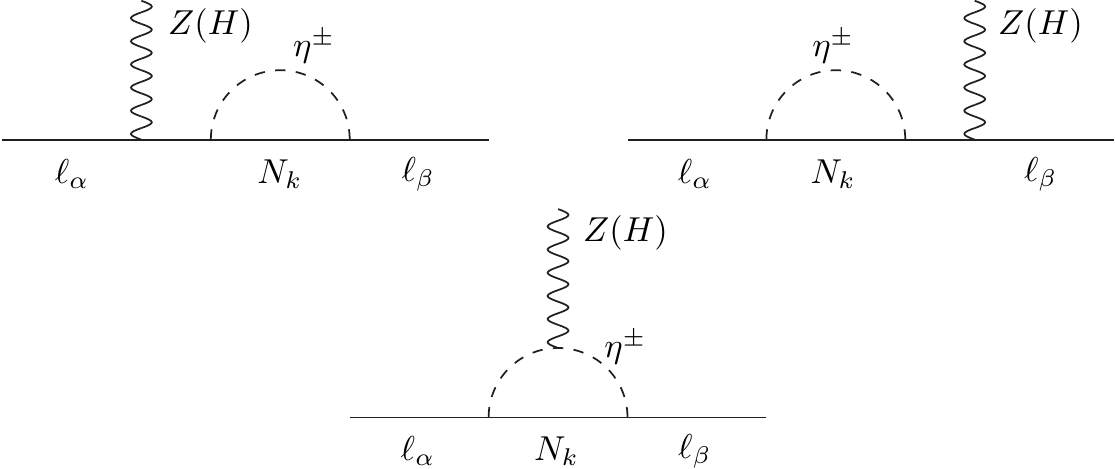}

\end{center}
\caption{Feynman diagrams representing the decays
$Z,H\to\ell_\alpha\ell_\beta$. In these diagrams, wavy line corresponds to
either $Z$ gauge boson or Higgs boson.}
\end{figure}
The amplitudes from the individual diagrams of Fig. 1 can have divergences.
But the sum of the amplitudes from
the diagrams of Fig. 1 is finite. For computing the amplitudes from the
diagrams of Fig. 1, we have followed the work of Ref. \cite{gl}.
In the individual diagrams of Fig. 1, we assign the momentum $p$ to the incoming
$Z$ or $H$. We assign momentum $p_1$ and $p_2$ to the outgoing charged
leptons $\ell_\alpha$ and $\ell_\beta$, respectively. In the
next two subsections, we present analytic results for the branching
ratios of $Z,H\to\ell_\alpha\ell_\beta$.

\subsection{Branching ratios of $Z\to\ell_\alpha\ell_\beta$}

In all the diagrams of Fig. 1, we can see propagators due to $\eta^\pm$
and $N_k$. Hence, it is convenient to define the following quantities
\begin{equation}
D_k=q^2-M_k^2,\quad D_{1\eta}=(q+p_1)^2-m_{\eta^\pm}^2,\quad
D_{2\eta}=(q-p_2)^2-m_{\eta^\pm}^2
\end{equation}
Here, $q$ is a 4-momentum.
While computing the amplitudes from the diagrams of Fig. 1, one come
across the following integrals \cite{pv}, through which we define the quantities
$b_{1,2}^k$, $c_{1,2}^k$, $d_{1,2}^k$, $f^k$ and $u^k$.
\begin{eqnarray}
&&\int\frac{d^dq}{(2\pi)^d}\frac{q^\mu}{D_kD_{1\eta}}=-b_1^kp_1^\mu,\quad
\int\frac{d^dq}{(2\pi)^d}\frac{q^\mu}{D_kD_{2\eta}}=b_2^kp_2^\mu,
\nonumber \\
&&\int\frac{d^dq}{(2\pi)^d}\frac{q^\mu}{D_kD_{1\eta}D_{2\eta}}=
-c_1^kp_1^\mu+c_2^kp_2^\mu,
\nonumber \\
&&\int\frac{d^dq}{(2\pi)^d}\frac{q^\mu q^\nu}{D_kD_{1\eta}D_{2\eta}}=
d_1^kp_1^\mu p_1^\nu+d_2^kp_2^\mu p_2^\nu-f^k(p_1^\mu p_2^\nu+p_2^\mu p_1^\nu)
+u^kg^{\mu\nu}
\label{int}
\end{eqnarray}
The above integrals are expressed in $d$-dimensions and at the end of the
calculations we take $d\to4$.
From these integrals, we can notice that $b_{1,2}^k$ and $u^k$ are
divergent quantities. On the other hand, $c_{1,2}^k$, $d_{1,2}^k$ and $f^k$
are finite.
Using the integrals of Eq. (\ref{int}), one can obtain the following relations
\begin{eqnarray}
&&b_1^k-b_2^k=(d_1^k-d_2^k)m_Z^2+(\kappa_1^k+\kappa_2^k)
(m_{\ell_\alpha}^2-m_{\ell_\beta}^2),
\label{b1-b2}
\\
&&m_{\ell_\alpha}^2b_1^k-m_{\ell_\beta}^2b_2^k=
(m_{\ell_\alpha}^2d_1^k-m_{\ell_\beta}^2d_2^k)m_Z^2
+(m_{\ell_\alpha}^2-m_{\ell_\beta}^2)[2u^k-f^km_Z^2+
\kappa_1^km_{\ell_\alpha}^2+\kappa_2^km_{\ell_\beta}^2],
\nonumber \\
\label{mb1-mb2}
\\
&&\kappa_1^k=d_1^k+f^k-c_1^k,\quad\kappa_2^k=d_2^k+f^k-c_2^k
\end{eqnarray}
Here, $m_Z$ is the mass of $Z$ gauge boson.

All the diagrams in Fig. 1 give divergent amplitudes for the case of
$Z\to\ell_\alpha\ell_\beta$. However, one can notice that the sum of the
amplitudes from these diagrams is
finite, after using Eqs. (\ref{b1-b2}) and (\ref{mb1-mb2}).
For the decay $Z\to\ell^+_\alpha\ell^-_\beta$, we have found the
total amplitude from the diagrams of Fig. 1 as
\begin{eqnarray}
-i{\cal M}_Z&=&\bar{u}(p_2)[A_1^L\gamma^\mu P_L+A_1^R\gamma^\mu P_R
+A_2^Li\sigma^{\mu\nu}p_\nu P_L+A_2^Ri\sigma^{\mu\nu}p_\nu P_R]v(p_1)
\epsilon_\mu(p),
\nonumber \\
&&P_{L(R)}=\frac{1\mp\gamma_5}{2},\quad\sigma^{\mu\nu}=\frac{i}{2}
[\gamma^\mu,\gamma^\nu],
\nonumber \\
A_1^L&=&\sum_{k=1}^3\frac{g}{c_W}(s_W^2-\frac{1}{2})h_{\alpha k}^*h_{\beta k}
(d_Z^k-f_Z^k)m_Z^2,\quad
A_1^R=\sum_{k=1}^3\frac{g}{c_W}h_{\alpha k}^*h_{\beta k}\kappa_Z^k
m_{\ell_\alpha}m_{\ell_\beta},
\nonumber \\
A_2^L&=&\sum_{k=1}^3\frac{g}{c_W}(s_W^2-\frac{1}{2})h_{\alpha k}^*h_{\beta k}
\kappa_Z^km_{\ell_\beta},\quad
A_2^R=\sum_{k=1}^3\frac{g}{c_W}(s_W^2-\frac{1}{2})h_{\alpha k}^*h_{\beta k}
\kappa_Z^km_{\ell_\alpha},
\nonumber \\
d_Z^k&=&d_1^k=d_2^k=\frac{-i}{16\pi^2}\int_0^1dx\int_0^{1-x}dy
\frac{y^2}{-y(1-x-y)m_Z^2+xM_k^2+(1-x)m_{\eta^\pm}^2},
\nonumber \\
f_Z^k&=&\frac{-i}{16\pi^2}\int_0^1dx\int_0^{1-x}dy
\frac{y(1-x-y)}{-y(1-x-y)m_Z^2+xM_k^2+(1-x)m_{\eta^\pm}^2},
\nonumber \\
c_Z^k&=&c_1^k=c_2^k=\frac{-i}{16\pi^2}\int_0^1dx\int_0^{1-x}dy
\frac{y}{-y(1-x-y)m_Z^2+xM_k^2+(1-x)m_{\eta^\pm}^2},
\nonumber \\
\kappa_Z^k&=&\kappa_1^k=\kappa_2^k=d_Z^k+f_Z^k-c_Z^k.
\end{eqnarray}
Here, $s_W(c_W)=\sin\theta_W(\cos\theta_W)$, where $\theta_W$ is the
weak-mixing angle. $g$ is the coupling strength of $SU(2)$ gauge group of
the standard model.
From the above amplitude, notice that, except $A_1^L$, rest of the form
factors of it are proportional to charged lepton masses. Since
$\frac{m_{\ell_\alpha}^2}{m_Z^2}\ll1$, the form factors $A_1^R$ and
$A_2^{L,R}$ give subleading contributions to the branching ratio of
$Z\to\ell^+_\alpha\ell^-_\beta$. As a result of this, the leading
contribution to the branching ratio of $Z\to\ell_\alpha\ell_\beta$ is
found to be
\begin{eqnarray}
{\rm Br}(Z\to\ell_\alpha\ell_\beta)&=&\frac{
\Gamma(Z\to\ell^+_\alpha\ell^-_\beta)+\Gamma(Z\to\ell^-_\alpha\ell^+_\beta)}
{\Gamma_Z}
\nonumber \\
&=&\left(\frac{g}{c_W}\right)^2\left(s_W^2-\frac{1}{2}\right)^2
\frac{m_Z^5}{12\Gamma_Z}
\left|\sum_{k=1}^3h_{\alpha k}^*h_{\beta k}(d_Z^k-f_Z^k)\right|^2
\label{brz}
\end{eqnarray}
Here, $\Gamma_Z$ is the total decay width of $Z$ gauge boson. In our
numerical analysis, which is presented in the next section, we have taken
$\Gamma_Z=$ 2.4952 GeV \cite{pdg}.

\subsection{Branching ratios of $H\to\ell_\alpha\ell_\beta$}

While computing the amplitude for $H\to\ell_\alpha\ell_\beta$, we can define
the integrals of Eq. (\ref{int}). Moreover, the relations in Eqs. (\ref{b1-b2})
and (\ref{mb1-mb2}) are also valid in this case after replacing $m_Z^2$ with
$m_H^2$ in these equations. Now, for the case of $H\to\ell_\alpha\ell_\beta$,
the top two diagrams of Fig. 1 give divergent amplitudes, whereas, the
bottom diagram of this figure give finite contribution. Hence, only the
analog of Eq. (\ref{b1-b2}) is
sufficient to see the cancellation of divergences between the top two diagrams
of Fig. 1. Now, after summing the amplitudes from the diagrams of Fig. 1,
for the decay $H\to\ell^+_\alpha\ell^-_\beta$, we have found the total
amplitude as
\begin{eqnarray}
i{\cal M}_H&=&\bar{u}(p_2)[A_H^LP_L+A_H^RP_R]v(p_1)
\nonumber \\
A_H^L&=&\sqrt{2}\sum_{k=1}^3h_{\alpha k}^*h_{\beta k}(\lambda_3c_H^k
+\frac{m_{\ell_\alpha}^2}{v^2}\kappa_H^k)vm_{\ell_\beta},
\nonumber \\
A_H^R&=&\sqrt{2}\sum_{k=1}^3h_{\alpha k}^*h_{\beta k}(\lambda_3c_H^k
+\frac{m_{\ell_\beta}^2}{v^2}\kappa_H^k)vm_{\ell_\alpha}
\end{eqnarray}
The expressions for $c_H^k$ and $\kappa_H^k$ are respectively same as that
for $c_Z^k$ and $\kappa_Z^k$, after replacing $m_Z^2$ with $m_H^2$ in these
expressions. The first term in $A_H^{L,R}$ is arising due to the bottom
diagram of Fig. 1. On the other hand, the top two diagrams of Fig. 1
contribute to
the second term in $A_H^{L,R}$. One can see that for $\lambda_3\sim1$, the
second term in $A_H^{L,R}$ gives negligibly small contribution. In our
numerical analysis, we consider $\lambda_3\sim1$. Hence, for a case like this,
the branching ratio for $H\to\ell_\alpha\ell_\beta$ is found to be
\begin{eqnarray}
{\rm Br}(H\to\ell_\alpha\ell_\beta)&=&\frac{
\Gamma(H\to\ell^+_\alpha\ell^-_\beta)+\Gamma(H\to\ell^-_\alpha\ell^+_\beta)}
{\Gamma_H}
\nonumber \\
&=&\frac{m_H}{4\pi\Gamma_H}(\lambda_3v)^2(m_{\ell_\alpha}^2+m_{\ell_\beta}^2)
\left|\sum_{k=1}^3h_{\alpha k}^*h_{\beta k}c_H^k\right|^2
\label{brh}
\end{eqnarray}
Here, $\Gamma_H$ is the total Higgs decay width.

In our numerical analysis,
which is presented in the next section, we have taken $m_H=$ 125.1 GeV
\cite{pdg} and $\Gamma_H=4.08\times 10^{-3}$ GeV \cite{yp}. This value of
$\Gamma_H$ is same as that for the Higgs boson of standard model. We have
taken this value for $\Gamma_H$ in order to simplify our numerical analysis.
The above mentioned value of $\Gamma_H$ has an implication that the
Higgs boson should not decay into $Z_2$-odd particles of the scotogenic
model. We comment further about this later.

\section{Numerical analysis}

From the analytic expressions given in the previous section, we can see that
the branching ratios of $Z,H\to\ell_\alpha\ell_\beta$ are proportional to the
Yukawa couplings $h_{\alpha k}$. The same Yukawa couplings also drive
neutrino masses which are described in Sec. 2. It is worth to explore the
correlation between neutrino oscillation observables and the branching ratios
of $Z,H\to\ell_\alpha\ell_\beta$. Here, our objective is
to fit the neutrino oscillation observables in the scotogenic model
in such a way that
the branching ratios for $Z,H\to\ell_\alpha\ell_\beta$ can become maximum
in this model. It
is explained in Sec. 2 that the above objective can be achieved by
taking $h_{\alpha k}\sim1$ and $\Lambda_k$ very small. Below we describe
the procedure in order to achieve this objective.

The neutrino oscillation observables can be
explained in the scotogenic model by parametrizing the Yukawa couplings as
given in Eq. (\ref{cipara}). In this equation, $R$ is an orthogonal matrix,
whose elements can have a magnitude of ${\cal O}(1)$. To simplify
our numerical analysis we take $R$ to be a unit matrix. In such a case we get
\begin{equation}
h=U^*_{PMNS}\cdot{\rm diag}\left(\sqrt{\frac{m_1}{\Lambda_1}},
\sqrt{\frac{m_2}{\Lambda_2}},\sqrt{\frac{m_3}{\Lambda_3}}\right)
\label{h}
\end{equation}
In our analysis we have parametrized $U_{PMNS}$ as \cite{pdg}
\begin{equation}
U_{PMNS} = \left(\begin{array}{ccc}
c_{12}c_{13} & s_{12}c_{13} & s_{13}e^{-i\delta_{CP}} \\
-s_{12}c_{23}-c_{12}s_{23}s_{13}e^{i\delta_{CP}} &
c_{12}c_{23}-s_{12}s_{23}s_{13}e^{i\delta_{CP}} & s_{23}c_{13} \\
s_{12}s_{23}-c_{12}c_{23}s_{13}e^{i\delta_{CP}} &
-c_{12}s_{23}-s_{12}c_{23}s_{13}e^{i\delta_{CP}} & c_{23}c_{13}
\end{array}\right)
\end{equation}
Here, $c_{ij}=\cos\theta_{ij}$, $s_{ij}=\sin\theta_{ij}$ and $\delta_{CP}$
is the $CP$ violating Dirac phase. We have taken Majorana phases to be
zero in $U_{PMNS}$. Shortly below we describe the numerical values
for neutrino masses and mixing angles. Using these values, we can see that
the elements of $U_{PMNS}$ can have a magnitude of ${\cal O}(1)$. Hence, we
need to make $\frac{m_k}{\Lambda_k}\sim1$ for $k=1,2,3$ in order to get
$h_{\alpha k}\sim1$. Since neutrino mass eigenvalues $m_k$ are very small,
$\Lambda_k$ should be proportionately small in order to achieve
$h_{\alpha k}\sim1$. It is described in Sec. 2 that $\Lambda_k$ can be
made very small by suppressing the $\lambda_5$ parameter.

%The Yukawa couplings in Eq. (\ref{h}) depend on neutrino masses and
%mixing angles. Below we describe the numerical values for these observables,
%which we use in our analysis.
From the global fits to neutrino oscillation
data the following mass-square differences among the neutrino fields are
found \cite{osci}.
\begin{equation}
m_s^2=m_2^2-m_1^2=7.5\times10^{-5}~{\rm eV}^2,\quad
m_a^2=\left\{\begin{array}{c}
m_3^2-m_1^2=2.55\times10^{-3}~{\rm eV}^2~~{\rm (NO)}\\
m_1^2-m_3^2=2.45\times10^{-3}~{\rm eV}^2~~{\rm (IO)}
\end{array}\right..
\label{msq}
\end{equation}
Here, NO(IO) represents normal(inverted) ordering. In the above equation
we have given the best fit values. In order to fit the
above mass-square differences, we take the neutrino mass eigenvalues as
\begin{eqnarray}
&& {\rm NO}:\quad m_1=0.1m_s,\quad m_2=\sqrt{m_s^2+m_1^2},\quad
m_3=\sqrt{m_a^2+m_1^2}.
\nonumber \\
&& {\rm IO}:\quad m_3=0.1m_s,\quad m_1=\sqrt{m_a^2+m_3^2},\quad
m_2=\sqrt{m_s^2+m_1^2}.
\label{nmass}
\end{eqnarray}
The above neutrino mass eigenvalues satisfy the cosmological upper bound
on the sum of neutrino masses, which is 0.12 eV \cite{cosmo}. Apart
from neutrino masses, neutrino
mixing angles are also found from the global fits to neutrino oscillation
data \cite{osci}. The best fit and 3$\sigma$ ranges for these variables
are given in Table 1.
\begin{table}[!h]
\centering
\begin{tabular}{|c|c c|} \hline
parameter & best fit & 3$\sigma$ range \\\hline
$\sin^2\theta_{12}/10^{-1}$ & 3.18 & 2.71 - 3.69 \\
$\sin^2\theta_{13}/10^{-2}$ (NO) & 2.200 & 2.000 - 2.405 \\
$\sin^2\theta_{13}/10^{-2}$ (IO) & 2.225 & 2.018 - 2.424 \\
$\sin^2\theta_{23}/10^{-1}$ (NO) & 5.74 & 4.34 - 6.10 \\
$\sin^2\theta_{23}/10^{-1}$ (IO) & 5.78 & 4.33 - 6.08 \\
$\delta_{CP}/{\rm o}$ (NO) & 194 & 128 - 359 \\
$\delta_{CP}/{\rm o}$ (IO) & 284 & 200 - 353 \\\hline
\end{tabular}
\caption{Best fit and 3$\sigma$ ranges for the neutrino mixing angles
and $CP$ violating Dirac phase, which are obtained from the global
fits to neutrino oscillation data \cite{osci}.}
\end{table}

In the next two subsections, we present numerical results on the branching
ratios of $Z,H\to\ell_\alpha\ell_\beta$. From the analytic expressions
given in the previous section, we can see that the above mentioned branching
ratios can become maximum for large values of Yukawa couplings and $\lambda_3$
parameter. In order to satisfy the perturbativity limits on these variables,
we apply the following constraints on the Yukawa couplings and the
$\lambda$ parameters of the scotogenic model.
\begin{equation}
|h_{\alpha k}|\leq\sqrt{4\pi},\quad|\lambda_i|\leq4\pi
\label{per}
\end{equation}

\subsection{$Z\to\ell_\alpha\ell_\beta$}

As explained previously, to satisfy perturbativity limit, $|h_{\alpha k}|$
can be as large as $\sqrt{4\pi}$. Since the magnitude of the elements of
$U_{PMNS}$ are less than about one, from Eq. (\ref{h}) we can see that
$\frac{m_k}{\Lambda_k}$ can be as large as 4$\pi$ in order to satisfy
the above mentioned perturbativity limit. $\Lambda_k$ depends on $M_k$, $m_0$
and $\lambda_5$. We have plotted $\frac{m_k}{\Lambda_k}$ versus $\lambda_5$
in Fig. 2.
\begin{figure}[!h]
\begin{center}

\includegraphics[width=3.0in]{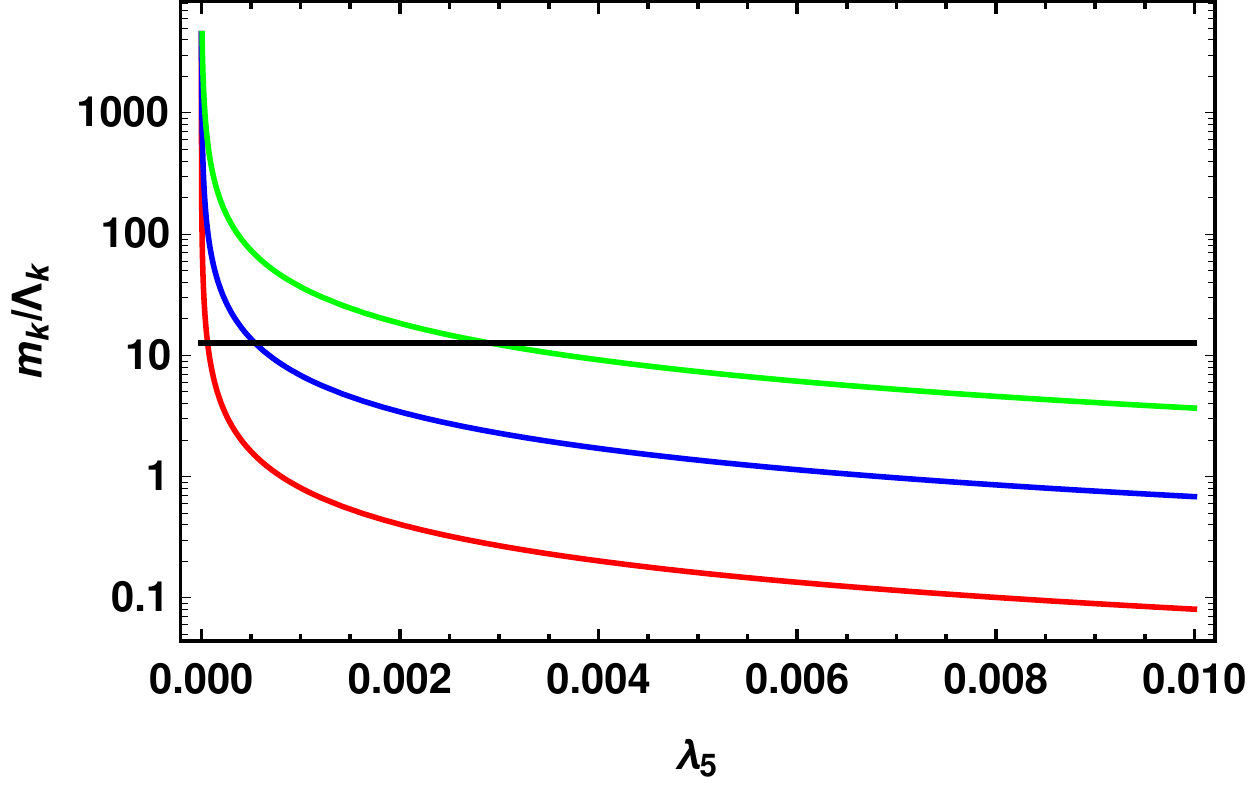}
\includegraphics[width=3.0in]{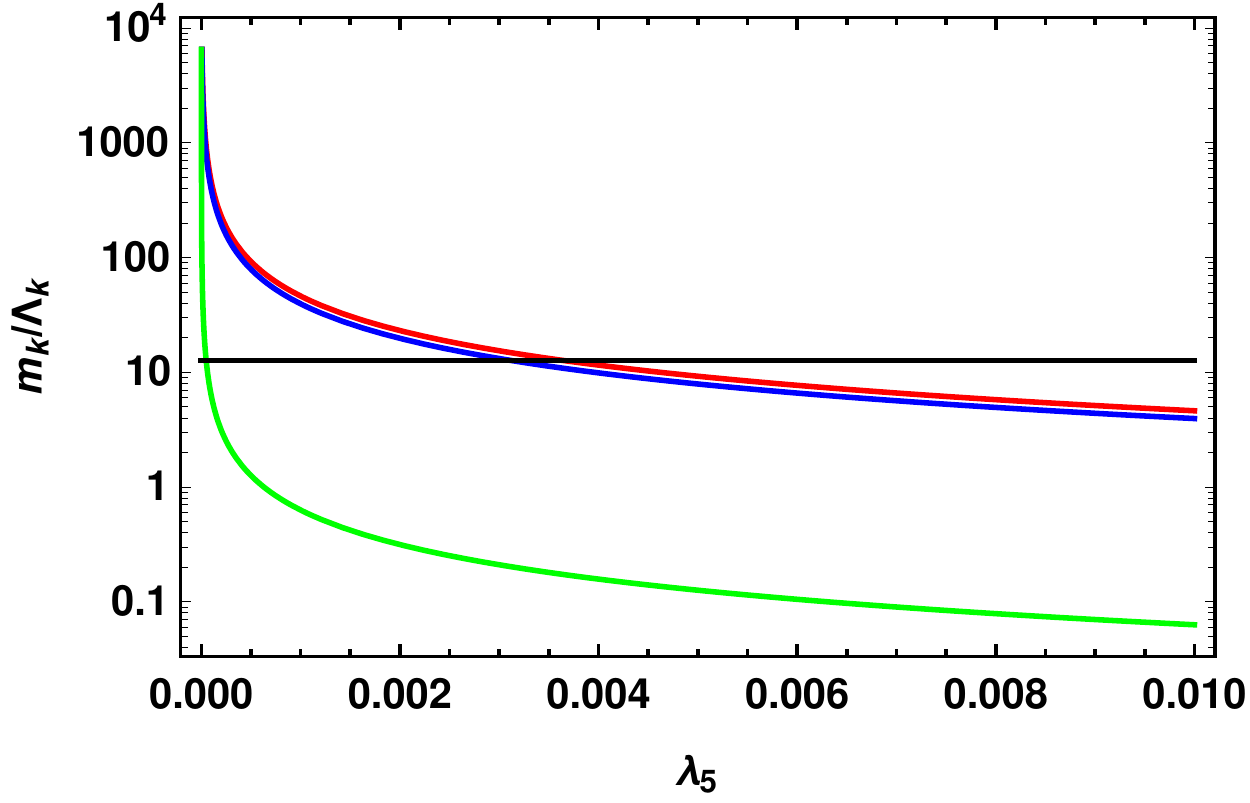}

\end{center}
\caption{Plots between $\frac{m_k}{\Lambda_k}$ and $\lambda_5$.
Red, blue and green lines are for $\frac{m_1}{\Lambda_1}$,
$\frac{m_2}{\Lambda_2}$ and $\frac{m_3}{\Lambda_3}$ respectively. Horizontal
line indicates the value 4$\pi$. Left- and right-hand side plots are
for NO and IO respectively. In both the plots, we have taken $m_0$ = 150 GeV,
$M_1$ = 100 GeV, $M_2=M_1+50$ GeV and $M_3=M_2+50$ GeV.}
\end{figure}
In these plots, we have chosen masses for right-handed neutrinos to be
between 100 to 200 GeV. The reason for such a choice is that, for these low
masses of right-handed neutrinos ${\rm Br}(Z\to\ell_\alpha\ell_\beta)$
can become maximum. Results related to ${\rm Br}(Z\to\ell_\alpha\ell_\beta)$
will be presented shortly later. In the plots of Fig. 2, for the case of
NO, all the lines are distinctly spaced because of the fact that the neutrino
masses are hierarchical in this case. On the other hand, the neutrino
mass eigenvalues $m_1$ and $m_2$ are nearly degenerate for the case of IO.
As a result of this, red and blue lines in the right-hand side plot of
Fig. 2 are close to each other. From this figure, we can see that
$\frac{m_k}{\Lambda_k}$ increases when $\lambda_5$ is decreasing. This follows
from the fact that in the limit $\lambda_5\to0$, $m_R^2$ and $m_I^2$ are
degenerate, and hence, $\Lambda_k$ becomes vanishingly small.
% From the
%above statement we can understand that when $\lambda_5$ is increasing
%the degeneracy between $m_R^2$ and $m_I^2$ decreases.
From Fig. 2, in the case of NO, for
$\lambda_5=3\times10^{-3}$ we get $\frac{m_3}{\Lambda_3}\approx4\pi$. Hence,
for $\lambda_5<3\times10^{-3}$ and for the values of $m_0,M_k$ taken in
Fig. 2, the perturbativity limit for Yukawa
couplings, which is given in Eq. (\ref{per}), can be violated. Similarly, from
the right-hand side plot of Fig. 2, we can see that the above mentioned
perturbativity limit can be violated for $\lambda_5<3.7\times10^{-3}$, in
the case of IO.

From Fig. 2, we have obtained the minimum value of $\lambda_5$ through which
the perturbativity limit on the Yukawa couplings can be satisfied. Using
this minimum value of $\lambda_5$ we have plotted branching ratios for
$Z\to\ell_\alpha\ell_\beta$ in Fig. 3 for the case of NO.
\begin{figure}[!h]
\begin{center}

\includegraphics[width=3.0in]{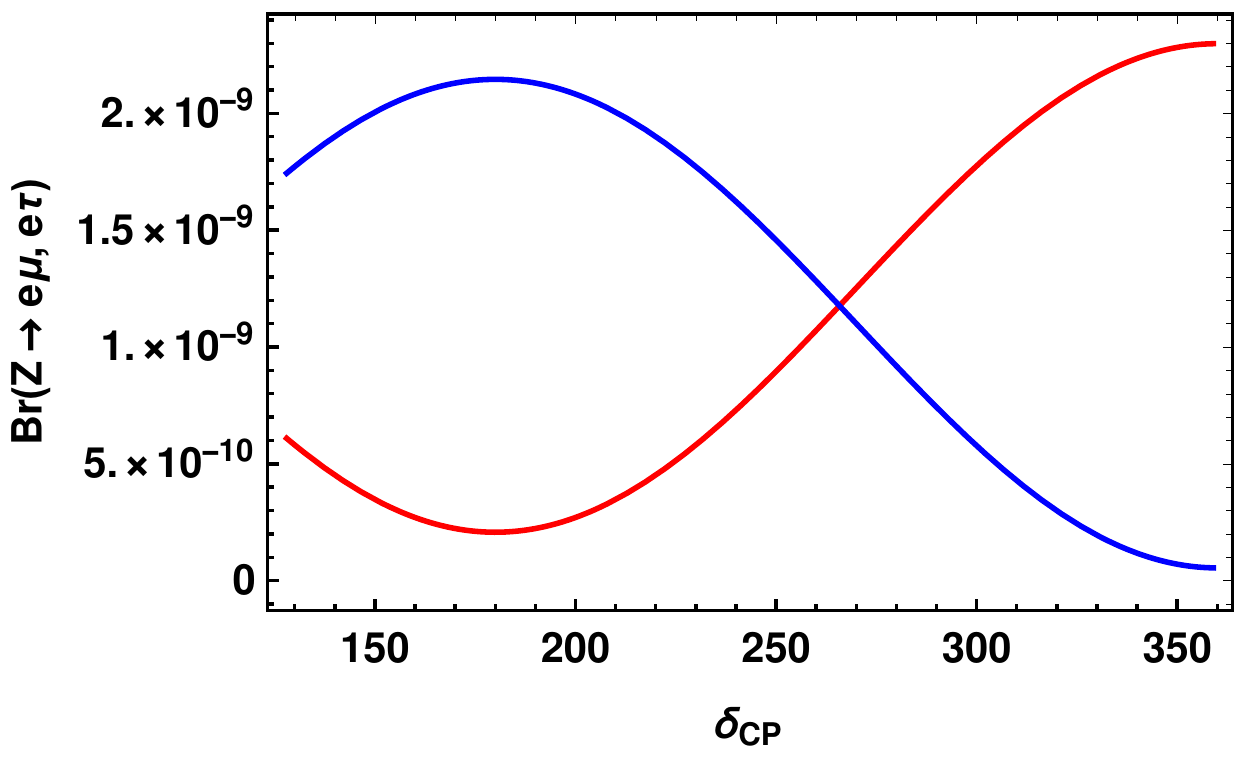}
\includegraphics[width=3.0in]{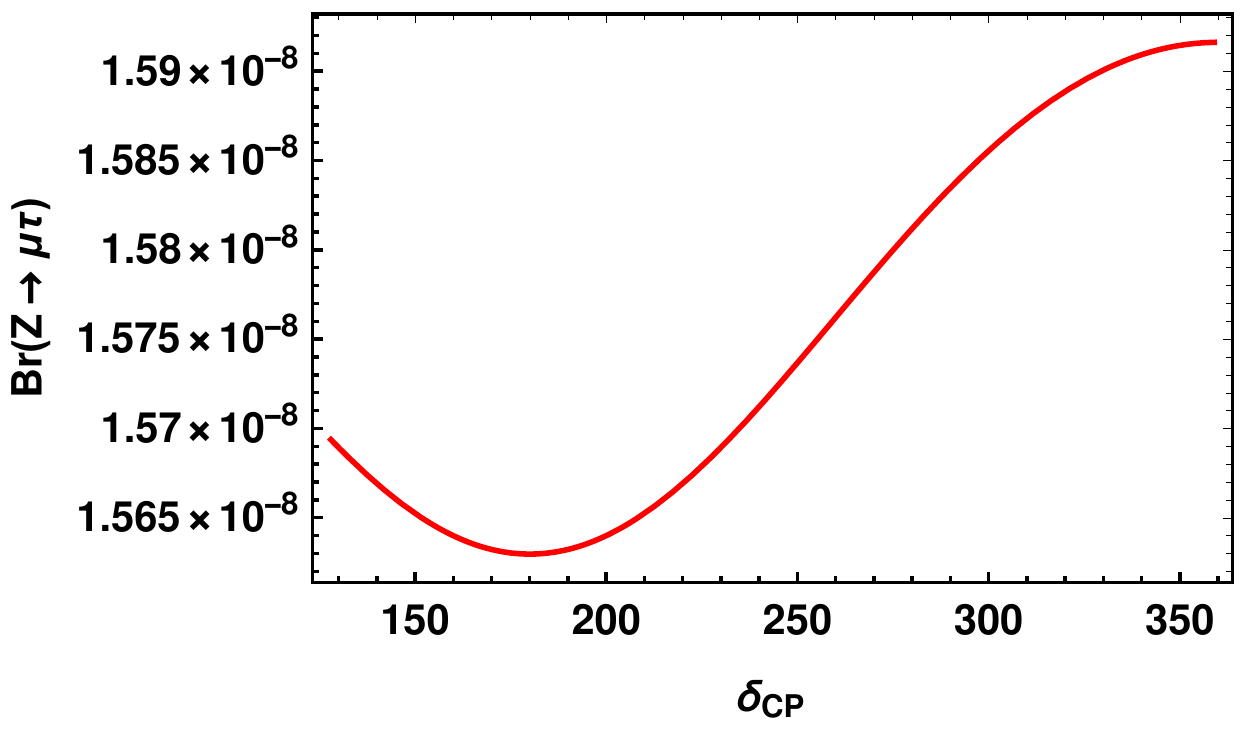}

\end{center}
\caption{Plots between ${\rm Br}(Z\to\ell_\alpha\ell_\beta)$ and $\delta_{CP}$
for the case of NO, without applying the constraints due to non-observation
of $\ell_\alpha\to\ell_\beta\gamma$.
Numerical values for neutrino masses are taken from Eq.
(\ref{nmass}). Neutrino mixing angles are taken to be the best fit values,
which are given in Table 1. In both the plots, we have taken
$\lambda_5=3\times10^{-3}$, $m_0$ = 150 GeV,
$m_{\eta^\pm}$ = 170 GeV, $M_1$ = 100 GeV, $M_2=M_1+50$ GeV and
$M_3=M_2+50$ GeV. In the left-hand side plot, red and blue lines are for
$e\mu$ and $e\tau$ modes respectively.}
\end{figure}
In the plots of this figure, we have taken $m_{\eta^\pm}$ to be as low as
170 GeV. One can understand that by increasing this value, branching
ratios for $Z\to\ell_\alpha\ell_\beta$ decreases. The plots in Fig. 3
are made after fitting to the neutrino oscillation observables in the
scotogenic model. We can see that the branching ratios for
$Z\to\ell_\alpha\ell_\beta$, in this model, can be as large as
$10^{-8}-10^{-9}$. These branching ratio values are lower than the current
experimental limits on them, which are given in Eq. (\ref{zexp}).
On the other hand, these values can be probed in the future FCC-ee collider,
which can be seen in Eq. (\ref{fzexp}). However, as will be described below,
the above mentioned branching ratio values will be suppressed, if constraints
due to non-observation of $\ell_\alpha\to\ell_\beta\gamma$ are applied.
We have also made the
analog plots of Fig. 3, for the case of IO, by taking
$\lambda_5=3.7\times10^{-3}$. We have found that, in the case of IO, the
branching ratios for $Z\to\ell_\alpha\ell_\beta$ are slightly higher
than that of plots in Fig. 3. But, otherwise, the shape of the curves for
${\rm Br}(Z\to\ell_\alpha\ell_\beta)$, in the case of IO, are same as that
of Fig. 3.

Regarding the shape of the curves in Fig. 3, we can notice that
the shapes of ${\rm Br}(Z\to e\mu)$ and ${\rm Br}(Z\to \mu\tau)$,
with respect to $\delta_{CP}$, are similar. On the other hand, the shapes of
${\rm Br}(Z\to e\mu)$ and ${\rm Br}(Z\to e\tau)$, with respect to
$\delta_{CP}$, are opposite to each other. We have found that the shapes of
the curves for ${\rm Br}(Z\to e\mu)$ and ${\rm Br}(Z\to e\tau)$, with respect
to $\delta_{CP}$, do not change by changing the values for neutrino mixing
angles. On the other hand, the shape of the curve for ${\rm Br}(Z\to \mu\tau)$, 
with respect to $\delta_{CP}$, changes with $s_{23}^2$. For $s_{23}^2>0.5$,
which is the case considered in Fig. 3, the shape of the curves for
${\rm Br}(Z\to e\mu)$ and ${\rm Br}(Z\to \mu\tau)$ are found to be similar.
In contrast to this, for $s_{23}^2<0.5$, the shape of the curve for
${\rm Br}(Z\to \mu\tau)$ is found to be similar to that of
${\rm Br}(Z\to e\tau)$. Whereas, for $s_{23}^2=0.5$, the shape of the
curve for ${\rm Br}(Z\to \mu\tau)$ has no resemblance with either to that
of ${\rm Br}(Z\to e\mu)$ and ${\rm Br}(Z\to e\tau)$. The shapes of the
above mentioned branching ratios with respect to $\delta_{CP}$ depend
on the Yukawa couplings, which in our case is given in Eq. (\ref{h}).
After using the form of these Yukawa couplings in the branching ratio
expressions of Eq. (\ref{brz}), one can understand the above described
shapes with respect to $\delta_{CP}$.

Plots in Fig. 3 are made for a minimum value of $\lambda_5$ for which
the Yukawa couplings can be close to a value of $\sqrt{4\pi}$. However,
the Yukawa couplings $h_{\alpha k}$ can also drive the decays
$\ell_\alpha\to\ell_\beta\gamma$, whose branching ratios in the
scotogenic model are as follows \cite{kms}.
\begin{eqnarray}
{\rm Br}(\ell_\alpha\to\ell_\beta\gamma)&=&\frac{3\alpha_{EW}}{64\pi G_F^2
m_{\eta^\pm}^4}\left|\sum_{k=1}^3h^*_{\alpha k}h_{\beta k}F_2\left(
\frac{M_k^2}{m_{\eta^\pm}^2}\right)\right|^2,
\nonumber \\
F_2(x)&=&\frac{1-6x+3x^2+2x^3-6x^2\ln x}{6(1-x)^4}
\label{brllg}
\end{eqnarray}
Here, $\alpha_{EW}$ and $G_F$ are fine-structure and Fermi constants,
respectively. The decays $\ell_\alpha\to\ell_\beta\gamma$ are not observed
in experiments. Hence, the branching ratios for these decays are constrained
as follows.
\begin{eqnarray}
{\rm Br}(\mu\to e\gamma)&<&4.2\times10^{-13}~\cite{meg},
\nonumber \\
{\rm Br}(\tau\to e\gamma)&<&3.3\times10^{-8}~\cite{babar},
\nonumber \\
{\rm Br}(\tau\to \mu\gamma)&<&4.4\times10^{-8}~\cite{babar}
\label{llgexp}
\end{eqnarray}
After comparing Eqs. (\ref{brz}) and (\ref{brllg}), we can see that the same
set of model parameters which determine ${\rm Br}(Z\to\ell_\alpha\ell_\beta)$
also determine ${\rm Br}(\ell_\alpha\to\ell_\beta\gamma)$. For the set of model
parameters taken in Fig. 3, we have found that the branching ratios for
$\ell_\alpha\to\ell_\beta\gamma$ exceed the experimental bounds of
Eq. (\ref{llgexp}). The reason for this is as follows. In the plots of Fig. 3,
the Yukawa couplings are close to $\sqrt{4\pi}$ and the masses of mediating
particles are between 100 to 200 GeV. For such large Yukawa couplings and
low masses, the branching ratios for $\ell_\alpha\to\ell_\beta\gamma$ are
quite large that they do not respect the bounds of Eq. (\ref{llgexp}).
Hence, the plots in Fig. 3 give us the maximum values that the branching
ratios of $Z\to\ell_\alpha\ell_\beta$ can reach in the scotogenic model,
without applying constraints due to non-observation of
$\ell_\alpha\to\ell_\beta\gamma$.
%But otherwise, the plots in Fig. 3 do not give the realistic values which
%can be probed in experiments.

Now, it is our interest to know the branching ratios of
$Z\to\ell_\alpha\ell_\beta$ after applying the constraints from
${\rm Br}(\ell_\alpha\to\ell_\beta\gamma)$. One can notice that
${\rm Br}(\ell_\alpha\to\ell_\beta\gamma)$ depends on Yukawa couplings,
masses of right-handed neutrinos and $\eta^\pm$. Hence, to satisfy
the bounds on ${\rm Br}(\ell_\alpha\to\ell_\beta\gamma)$, one has to
suppress Yukawa couplings and increase the masses for right-handed
neutrinos and $\eta^\pm$. The mass of $\eta^\pm$ can be written as
$m_{\eta^\pm}=\sqrt{m_0^2-\lambda_4 v^2}$. To satisfy the perturbativity
limit on $\lambda_4$, we choose $\lambda_4=-4\pi$. With this choice,
the mass of $\eta^\pm$ can take maximum value, for a fixed value of $m_0$.
Now, the Yukawa couplings depend on $m_0$, $\lambda_5$ and masses of
right-handed neutrinos, apart from neutrino oscillation observables.
Hence, for the above mentioned choice, ${\rm Br}(Z\to\ell_\alpha\ell_\beta)$
and ${\rm Br}(\ell_\alpha\to\ell_\beta\gamma)$ depend on
$m_0$, $\lambda_5$ and masses of
right-handed neutrinos, apart from neutrino oscillation observables.
In Fig. 4, we have plotted branching ratios of $Z\to\ell_\alpha\ell_\beta$
after applying the constraints from ${\rm Br}(\ell_\alpha\to\ell_\beta\gamma)$.
\begin{figure}[!h]
\begin{center}

\includegraphics[width=3.0in]{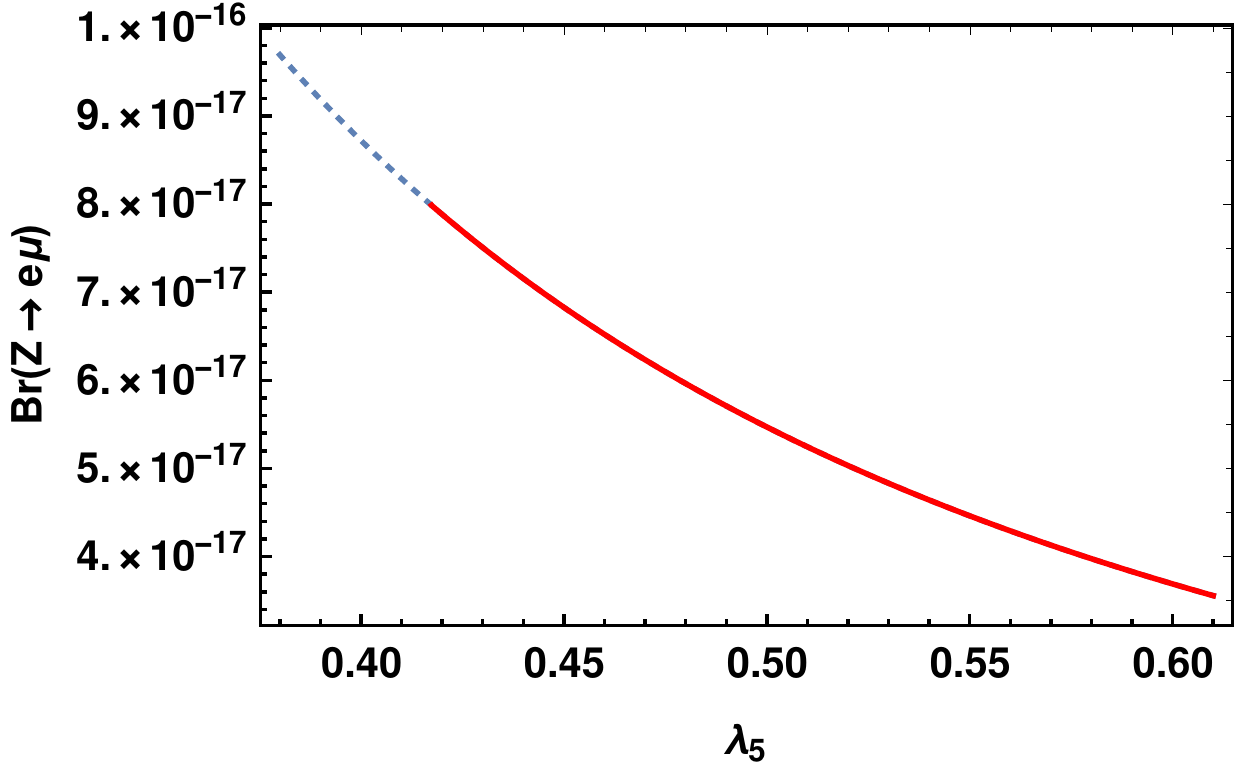}
\includegraphics[width=3.0in]{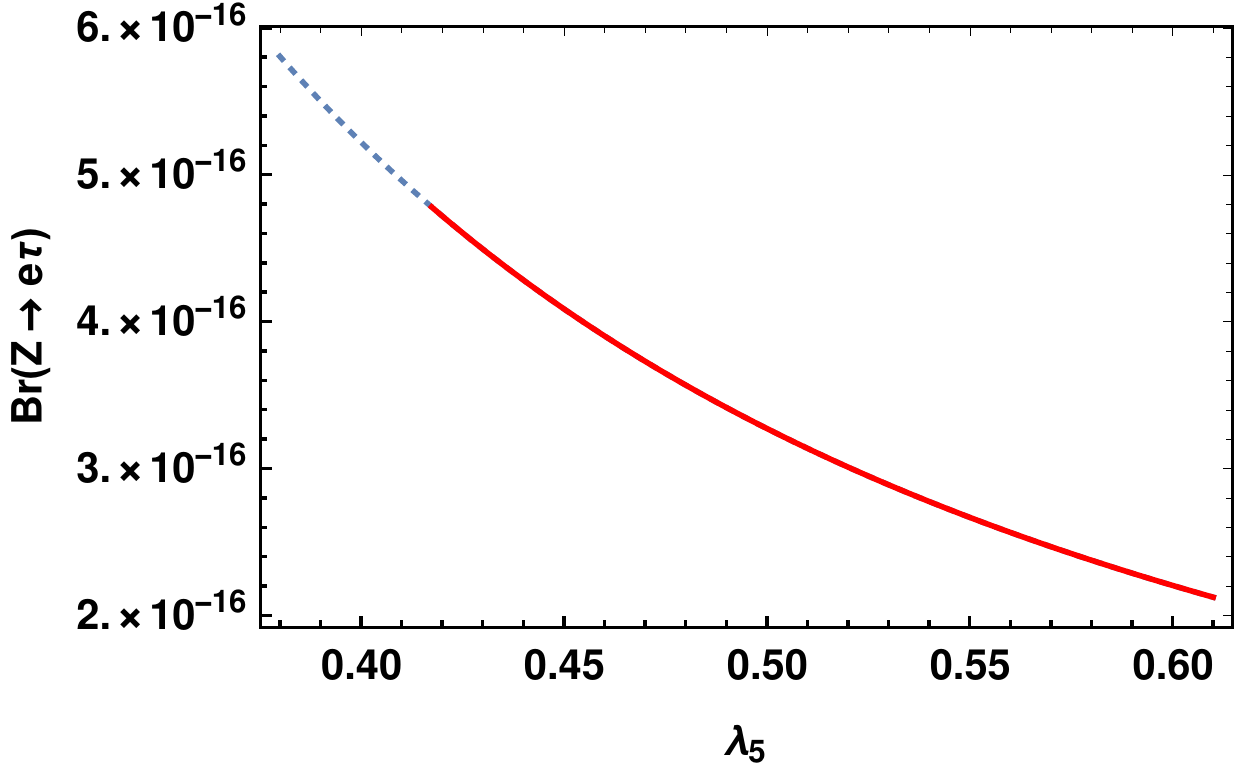}
\includegraphics[width=3.0in]{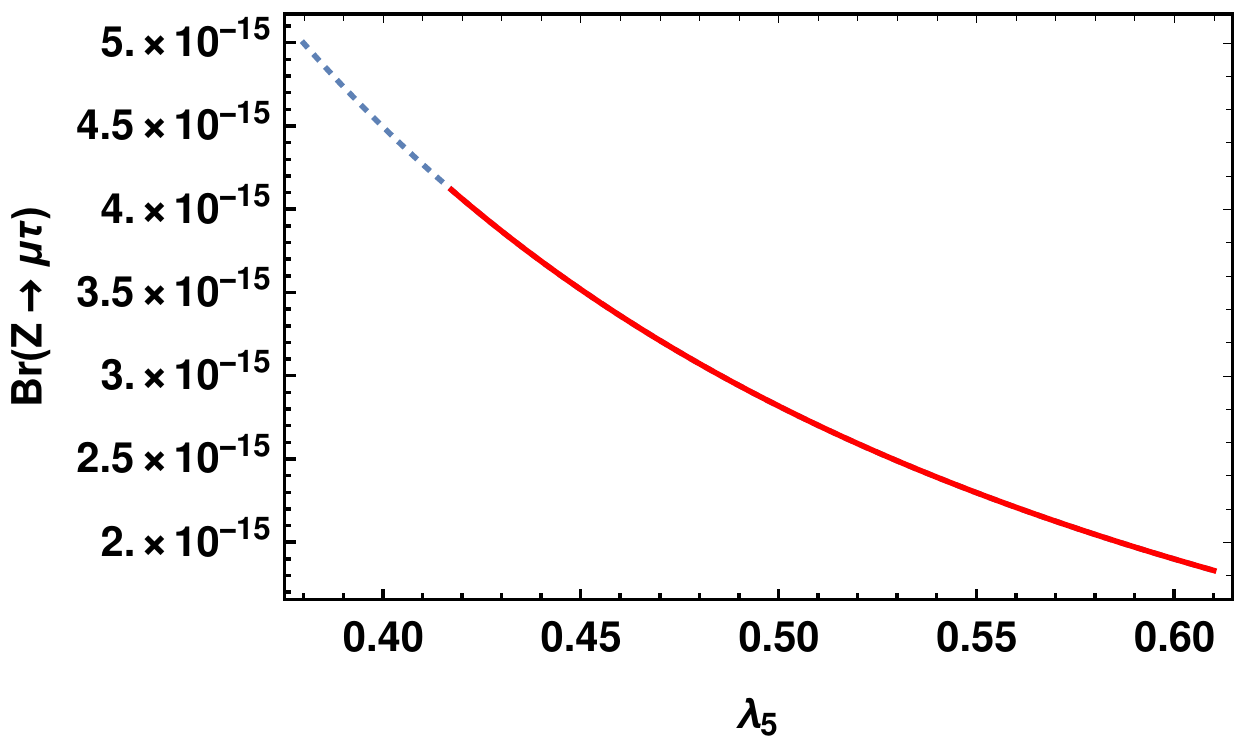}

\end{center}
\caption{Plots between ${\rm Br}(Z\to\ell_\alpha\ell_\beta)$ and $\lambda_5$
for the case of NO, after applying the constraints from ${\rm Br}(\ell_\alpha
\to\ell_\beta\gamma)$. In these plots, solid lines are allowed and dotted lines
are excluded by the constraints due to
${\rm Br}(\ell_\alpha\to\ell_\beta\gamma)$. Numerical values for
neutrino masses are taken from Eq.
(\ref{nmass}). Neutrino mixing angles and $\delta_{CP}$ are taken to be the
best fit values, which are given in Table 1. We have taken
$m_0$ = 150 GeV,
$m_{\eta^\pm}=\sqrt{m_0^2+4\pi v^2}$, $M_1$ = 1000 GeV, $M_2=M_1+100$ GeV and
$M_3=M_2+100$ GeV.}
\end{figure}

In Fig. 4, we have varied $\lambda_5$ up to 0.61. The reason for this
is explained below. For the parametric values of Fig. 4, we can see that
the lightest $Z_2$-odd particle in the scotogenic model is $\eta_I$. The
mass of $\eta_I$ decreases with $\lambda_5$. At $\lambda_5=0.61$, we get
$m_I\approx$ 63.5 GeV. Since the Higgs boson mass is 125.1 GeV, for
$\lambda_5>0.61$ there is a possibility that the Higgs can decay into a pair
of $\eta_I$. It is described in the previous section that the
total decay width for the Higgs boson in our analysis is taken to be
the same as that in the standard model. Hence, to avoid the above mentioned
decay, we have varied $\lambda_5$ up to 0.61 in Fig. 4.

From Fig. 4, we can see that the branching ratios for
$Z\to\ell_\alpha\ell_\beta$ vary in the range of $10^{-17}-10^{-15}$. These
values are suppressed by about $10^{-7}$ as compared that in Fig. 3.
The reason for
this suppression is due to the fact that the $\lambda_5$ and the masses of
right-handed neutrinos and $\eta^\pm$ are large as compared
to those in Fig. 3. As already stated before, the masses of right-handed
neutrinos and $\eta^\pm$ should be taken large, otherwise, the constraints
on ${\rm Br}(\ell_\alpha\to\ell_\beta\gamma)$ cannot be satisfied. The mass
of lightest right-handed neutrino in Fig. 4 is taken to be 1 TeV. We have
found that, for the case of $M_2=M_1+100$ GeV and $M_3=M_2+100$ GeV,
$M_1$ should be
at least around 500 GeV in order to satisfy the constraints from
${\rm Br}(\ell_\alpha\to\ell_\beta\gamma)$. However, in such a case,
the allowed range for $\lambda_5$ becomes narrower than that in Fig. 4 and
the allowed ranges for ${\rm Br}(Z\to\ell_\alpha\ell_\beta)$ are found to
be nearly same as that in Fig. 4. Although the right-handed neutrino masses
are taken
to be non-degenerate in Fig. 4, the plots in this figure do not vary much
with degenerate right-handed neutrinos of 1 TeV masses.
It is stated above that another reason for the suppression of
${\rm Br}(Z\to\ell_\alpha\ell_\beta)$ in Fig. 4 is due to the fact that
$\lambda_5$ is
large. This suppression is happening because Yukawa couplings reduce
with increasing $\lambda_5$. This fact can be understood with the plots of
Fig. 2 and also with Eq. (\ref{h}).

In the plots of Fig. 4, we have fixed $m_0$ to 150 GeV. By increasing this
value to 500 GeV, we have found that ${\rm Br}(Z\to\ell_\alpha\ell_\beta)$
reduces as compared to that in Fig. 4. This is happening because
$m_{\eta^\pm}$ increases. Another difference we have noticed is that, for
$m_0=$ 500 GeV and right-handed neutrino masses to be same as in Fig. 4,
the allowed range for $\lambda_5$ is found to be $\sim1.5-8.0$. This
is happening because, by increasing $m_0$, one has to increase $\lambda_5$
in order to suppress the Yukawa couplings and thereby satisfy the
constraints on ${\rm Br}(\ell_\alpha\to\ell_\beta\gamma)$.

We have plotted ${\rm Br}(Z\to\ell_\alpha\ell_\beta)$ for the case of IO,
which are presented in Fig. 5.
\begin{figure}[!h]
\begin{center}

\includegraphics[width=3.0in]{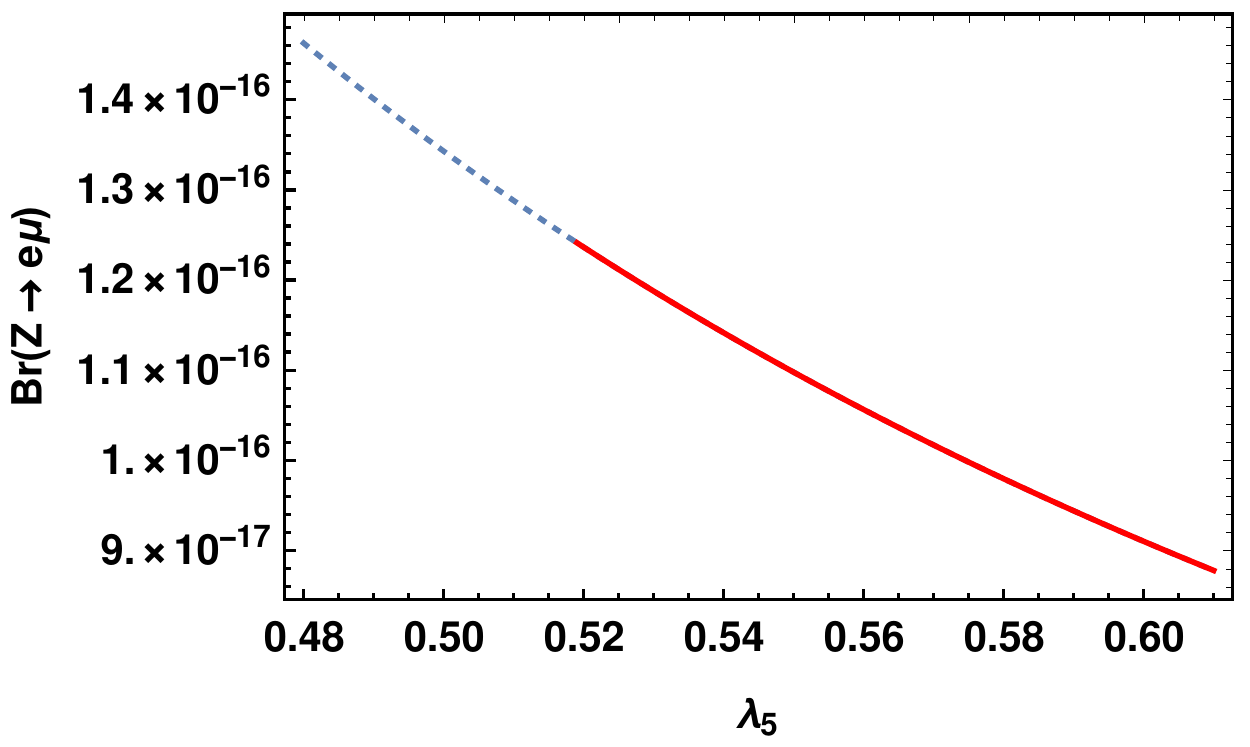}
\includegraphics[width=3.0in]{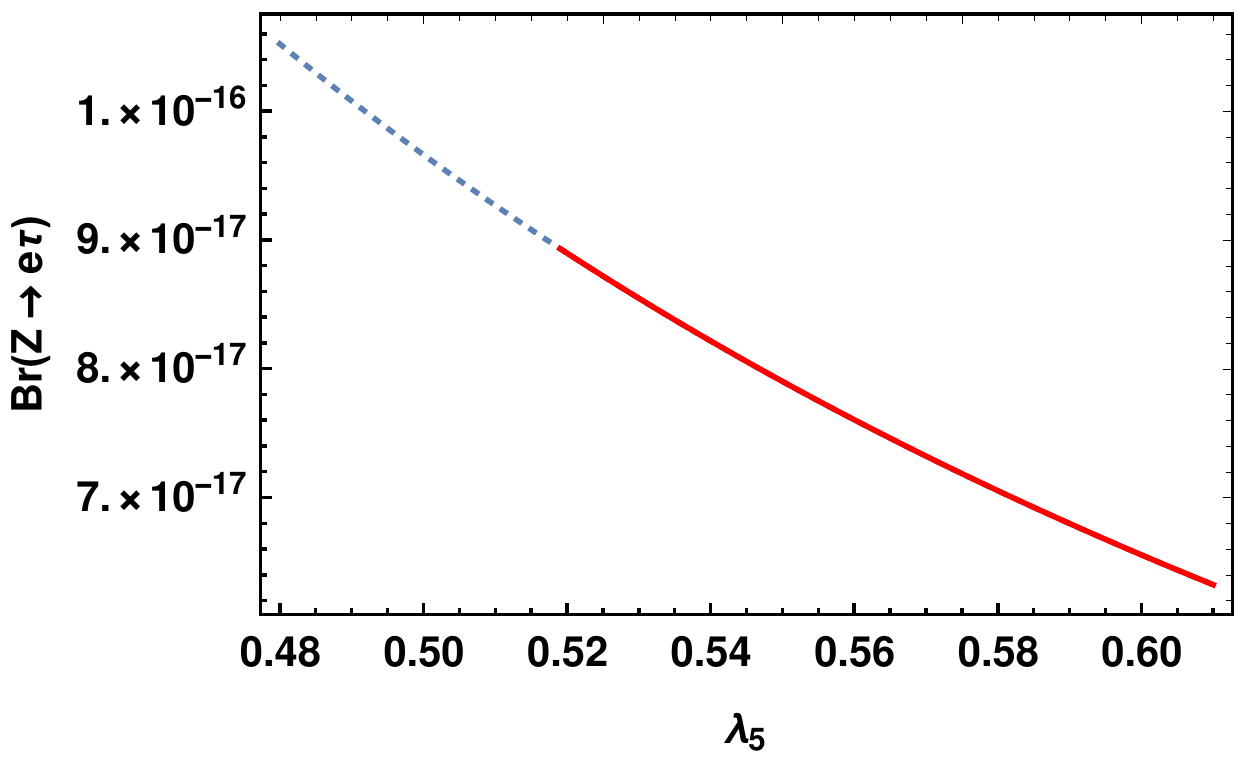}
\includegraphics[width=3.0in]{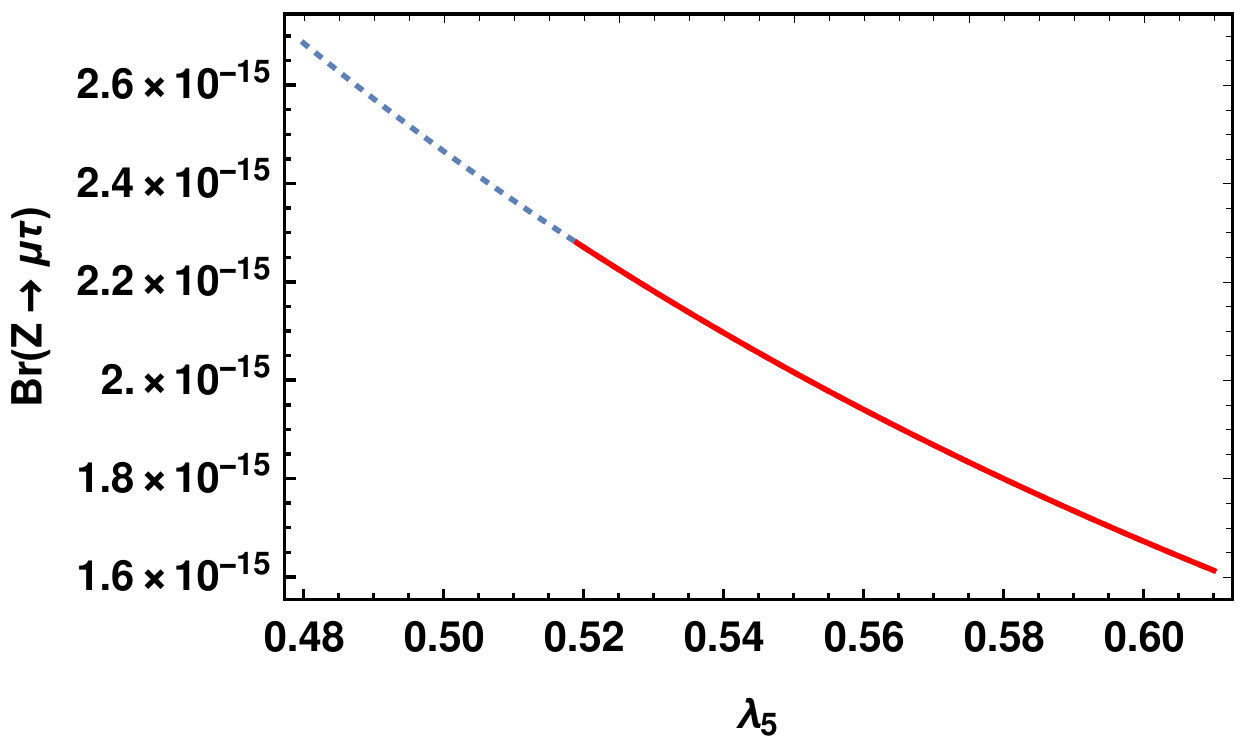}

\end{center}
\caption{Plots between ${\rm Br}(Z\to\ell_\alpha\ell_\beta)$ and $\lambda_5$
for the case of IO, after applying the constraints from ${\rm Br}(\ell_\alpha
\to\ell_\beta\gamma)$. In these plots, solid lines are allowed and dotted lines
are excluded by the constraints due to
${\rm Br}(\ell_\alpha\to\ell_\beta\gamma)$. Numerical values for
neutrino masses are taken from Eq.
(\ref{nmass}). Neutrino mixing angles and $\delta_{CP}$ are taken to be the
best fit values, which are given in Table 1. We have taken
$m_0$ = 150 GeV,
$m_{\eta^\pm}=\sqrt{m_0^2+4\pi v^2}$, $M_1$ = 2000 GeV, $M_2=M_1+100$ GeV and
$M_3=M_2+100$ GeV.}
\end{figure}
In this case, for $m_0=$ 150 GeV
we have found that $M_1$ should be at least around 1.7 TeV in order to
satisfy the constraints on ${\rm Br}(\ell_\alpha\to\ell_\beta\gamma)$.
As a result of this, in the plots of Fig. 5, we have taken $M_1=$ 2 TeV.
Comparing the plots of Figs. 4 and 5, we can conclude the following points.
In both the cases of NO and IO, ${\rm Br}(Z\to\mu\tau)$ is larger than
that for the other LFV decays of $Z$ gauge boson. In the case of NO,
${\rm Br}(Z\to e\mu)$ is one order less than ${\rm Br}(Z\to e\tau)$.
On the other hand, in the case of IO, ${\rm Br}(Z\to e\mu)$ is slightly
larger than ${\rm Br}(Z\to e\tau)$.

\subsection{$H\to\ell_\alpha\ell_\beta$}

In this subsection, we present numerical results on the branching ratios of
$H\to\ell_\alpha\ell_\beta$. After comparing Eqs. (\ref{brz}) and (\ref{brh}),
we can see that a common set of parameters determine both
${\rm Br}(Z\to\ell_\alpha\ell_\beta)$ and
${\rm Br}(H\to\ell_\alpha\ell_\beta)$. Apart from this common set of
parameters, $\lambda_3$ is an additional parameter which determine
${\rm Br}(H\to\ell_\alpha\ell_\beta)$. In our analysis, we have taken
$\lambda_3=4\pi$ in order to satisfy the perturbativity limit and also to
maximize ${\rm Br}(H\to\ell_\alpha\ell_\beta)$. Apart from the above mentioned
parameters, ${\rm Br}(H\to\ell_\alpha\ell_\beta)$ also depends on the
charged lepton masses. We have taken these masses to be the best fit values,
which are given in Ref. \cite{pdg}.

First we present the results on
${\rm Br}(H\to\ell_\alpha\ell_\beta)$ after fitting to the neutrino
oscillation observables, but without satisfying the constraints from
${\rm Br}(\ell_\alpha
\to\ell_\beta\gamma)$. These results are given in Fig. 6 for the case of NO.
\begin{figure}[!h]
\begin{center}

\includegraphics[width=3.0in]{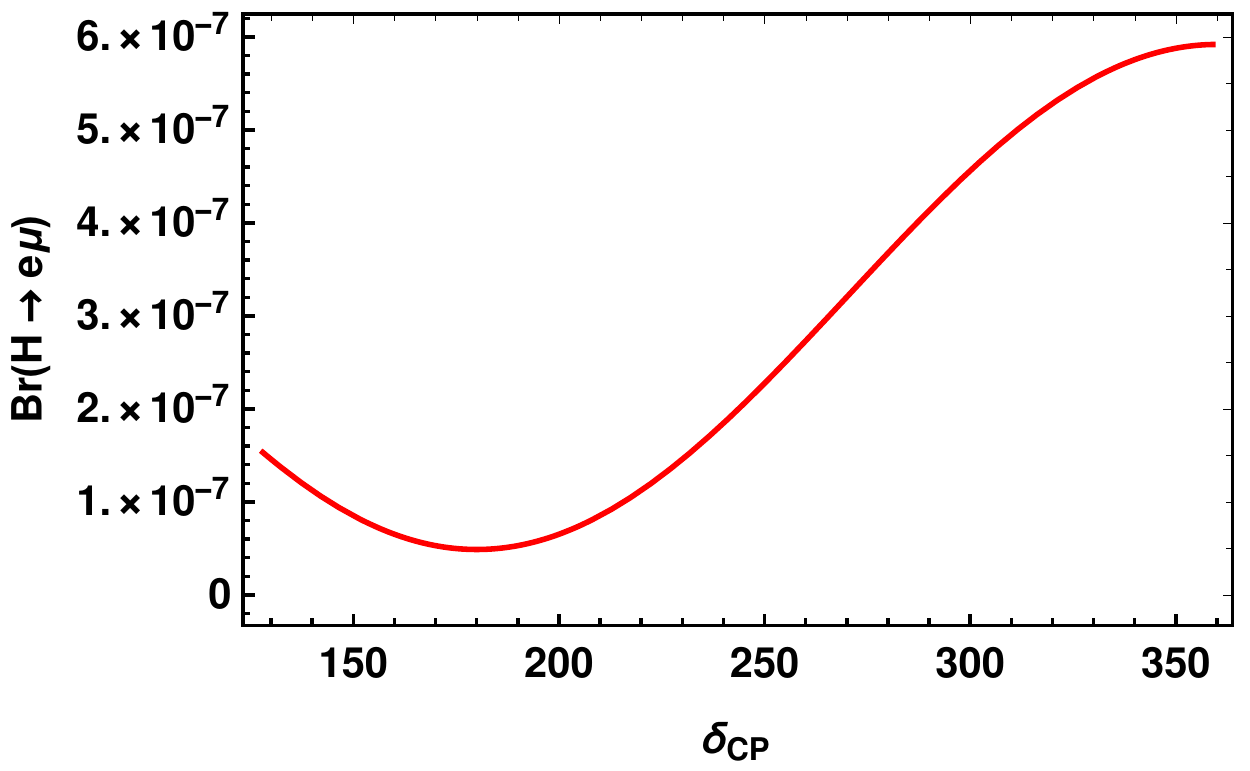}
\includegraphics[width=3.0in]{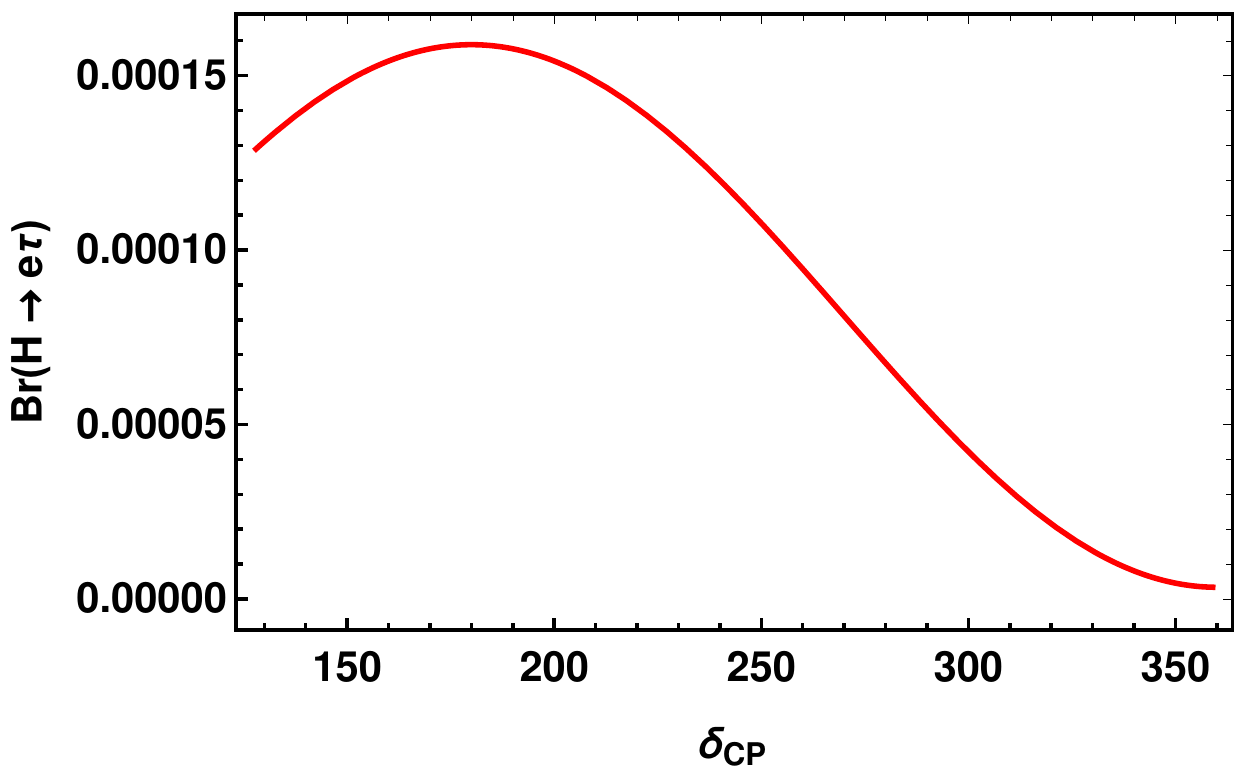}
\includegraphics[width=3.0in]{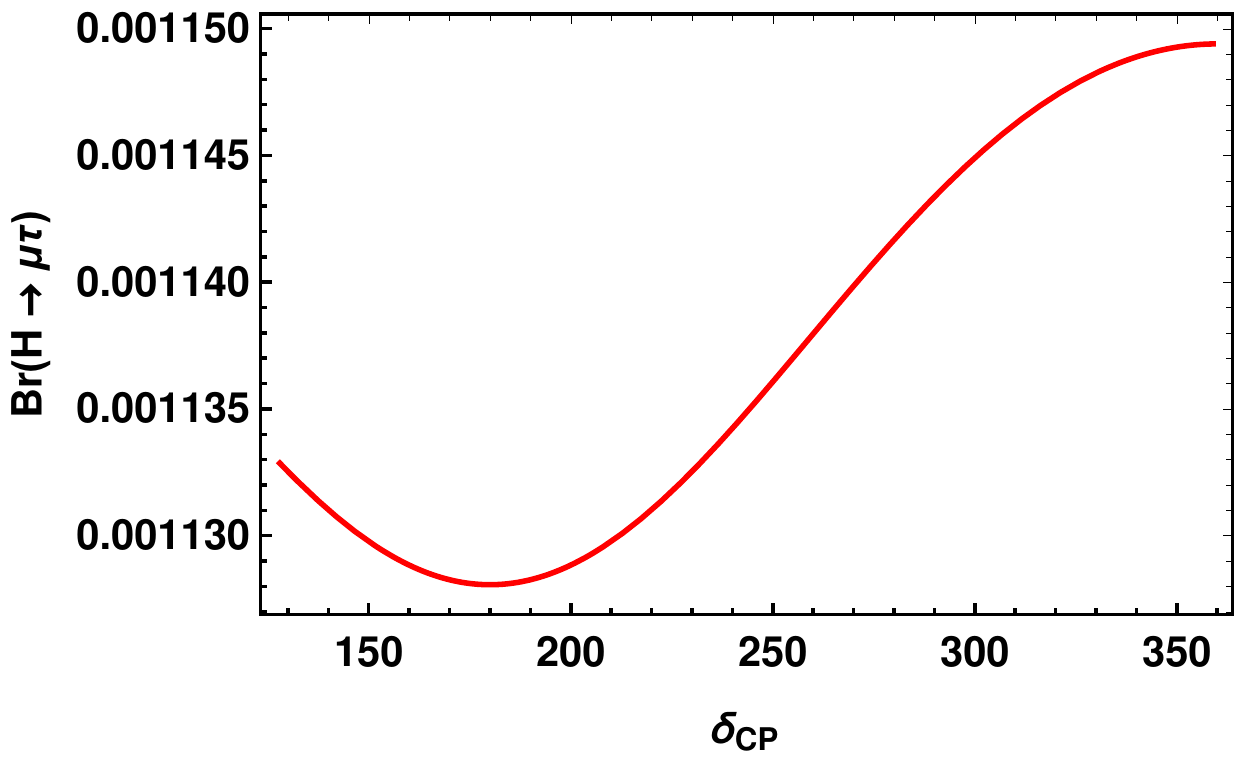}

\end{center}
\caption{Plots between ${\rm Br}(H\to\ell_\alpha\ell_\beta)$ and $\delta_{CP}$
for the case of NO, without applying the constraints from ${\rm Br}(\ell_\alpha
\to\ell_\beta\gamma)$. See the caption of Fig. 3, for parametric values and
neutrino oscillation observables, which are used in these plots.}
\end{figure}
One can compare the branching ratios in this figure with the current
limits on them, which are given in Eq. (\ref{hexp}). We can see that
the values for ${\rm Br}(H\to e\mu)$ and ${\rm Br}(H\to e\tau)$ from
this figure are marginally lower than the current experimental limits on them.
Whereas, the
values for ${\rm Br}(H\to \mu\tau)$ are just below the current experimental
limit on this. However, in the plots of Fig. 6, we have taken
$\lambda_5=3\times10^{-3}$ and the masses of right-handed neutrinos
and $\eta^\pm$ are chosen to be between 100 to 200 GeV. For this choice
of parameters, as already explained in the previous subsection, the
Yukawa couplings can be large, and hence, ${\rm Br}(H\to\ell_\alpha\ell_\beta)$
can become maximum. Plots in Fig. 6 are made for the case of NO. We have
plotted ${\rm Br}(H\to\ell_\alpha\ell_\beta)$ for the case of IO by
taking $\lambda_5=3.7\times10^{-3}$ and for the mass parameters which
are described above. In this case, we have found a slight enhancement
in the values of ${\rm Br}(H\to\ell_\alpha\ell_\beta)$ as compared to that
of Fig. 6. But otherwise, in the case of IO, the shape of the curves for
${\rm Br}(H\to\ell_\alpha\ell_\beta)$ are found to be the same as that in
Fig. 6.

In the plots of Fig. 6, constraints from
${\rm Br}(\ell_\alpha\to\ell_\beta\gamma)$
are not applied. After applying the constraints from
${\rm Br}(\ell_\alpha\to\ell_\beta\gamma)$, branching ratios for
$H\to\ell_\alpha\ell_\beta$ are given in Fig. 7 for the case of NO.
\begin{figure}[!h]
\begin{center}

\includegraphics[width=3.0in]{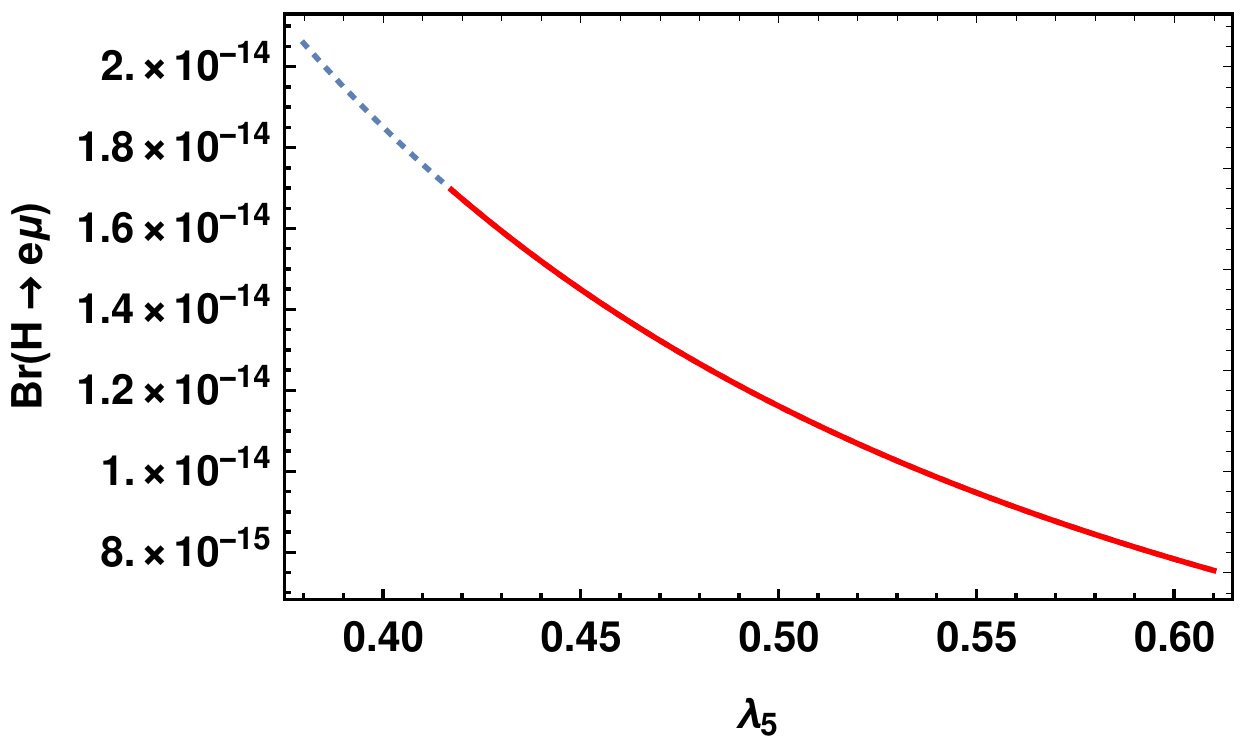}
\includegraphics[width=3.0in]{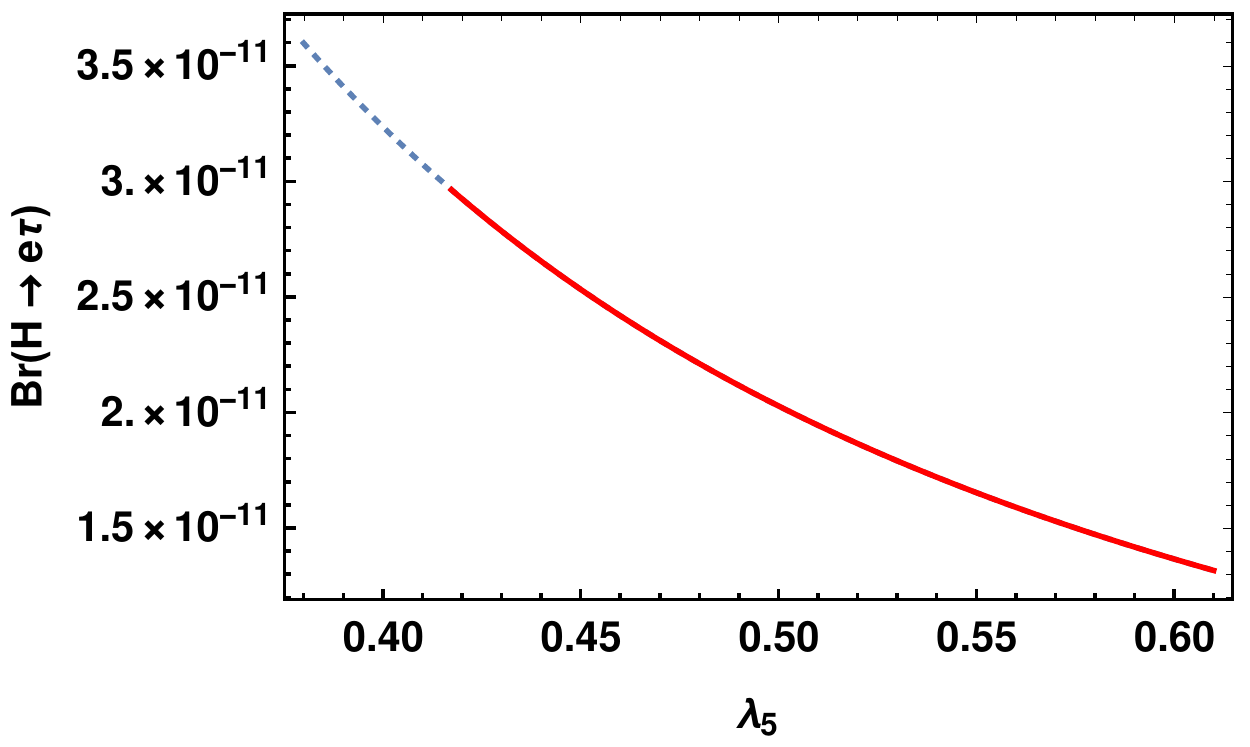}
\includegraphics[width=3.0in]{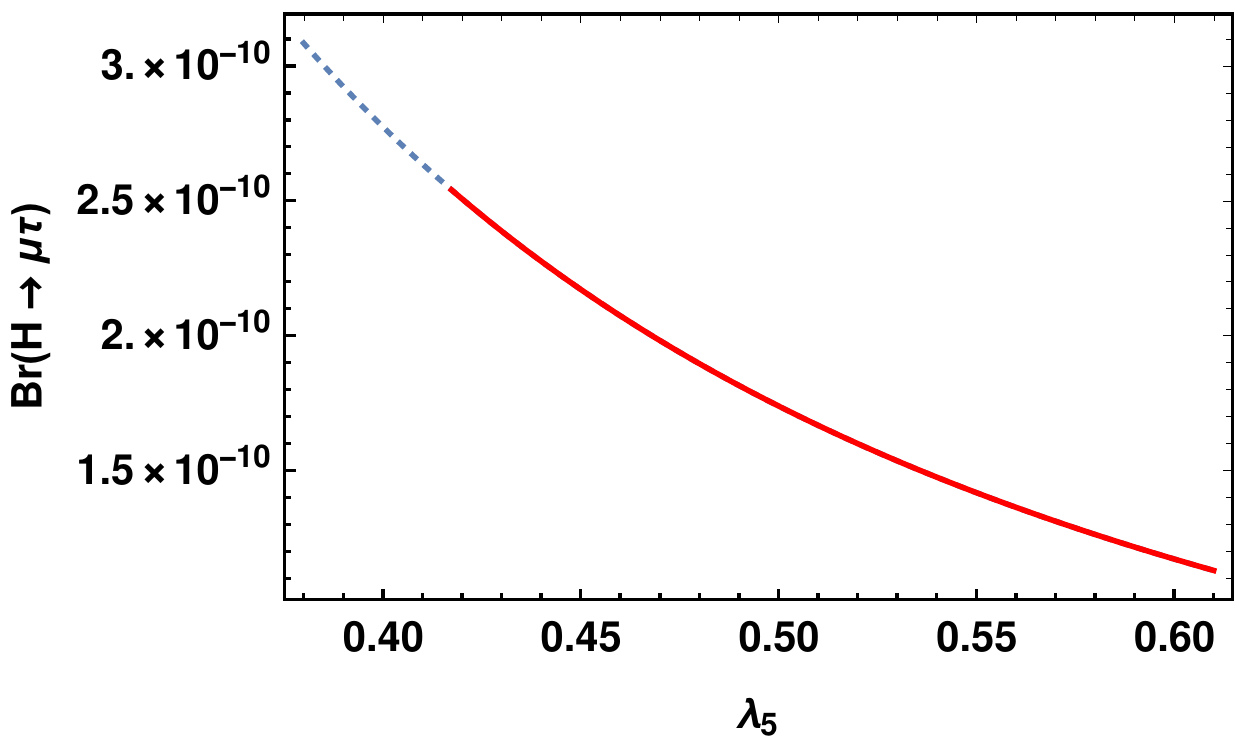}

\end{center}
\caption{Plots between ${\rm Br}(H\to\ell_\alpha\ell_\beta)$ and $\lambda_5$
for the case of NO, after applying the constraints from ${\rm Br}(\ell_\alpha
\to\ell_\beta\gamma)$. In these plots, solid lines are allowed and dotted lines
are excluded by the constraints due to
${\rm Br}(\ell_\alpha\to\ell_\beta\gamma)$.
See the caption of Fig. 4, for parametric values and
neutrino oscillation observables, which are used in these plots.}
\end{figure}
One can see that the branching ratios in this figure are suppressed by a
factor of about $10^{-7}$ as compared to that in Fig. 6. The reason for this
suppression, which can be understood from the reasoning's given around
Fig. 4, is due to the fact that $\lambda_5$ and masses of right-handed
neutrinos and $\eta^\pm$ are large as compared that in Fig. 6. The mass
of lightest right-handed neutrino is 1 TeV in Fig. 7. As already pointed around
Fig. 4, the value of $M_1$ should be at least around 500 GeV in order to
satisfy the constraints from ${\rm Br}(\ell_\alpha\to\ell_\beta\gamma)$
for the case of Fig. 7. Even with $M_1=$ 500 GeV, we have found the
allowed ranges for ${\rm Br}(H\to\ell_\alpha\ell_\beta)$ are nearly same
as that of Fig. 7. Although
the right-handed neutrino masses are non-degenerate in Fig. 7, with
degenerate right-handed neutrinos with masses of 1 TeV we have found that
the allowed ranges for ${\rm Br}(H\to\ell_\alpha\ell_\beta)$ are similar
to that in Fig. 7. In this figure, among the three LFV decays of
$H$, the branching ratios of $H$ into $\tau$ mode are large, since these
branching ratios are proportional to $m_\tau^2$.

We have plotted ${\rm Br}(H\to\ell_\alpha\ell_\beta)$, after applying
the constraints from ${\rm Br}(\ell_\alpha\to\ell_\beta\gamma)$,
for the case of IO.
These plots are given in Fig. 8.
\begin{figure}[!h]
\begin{center}

\includegraphics[width=3.0in]{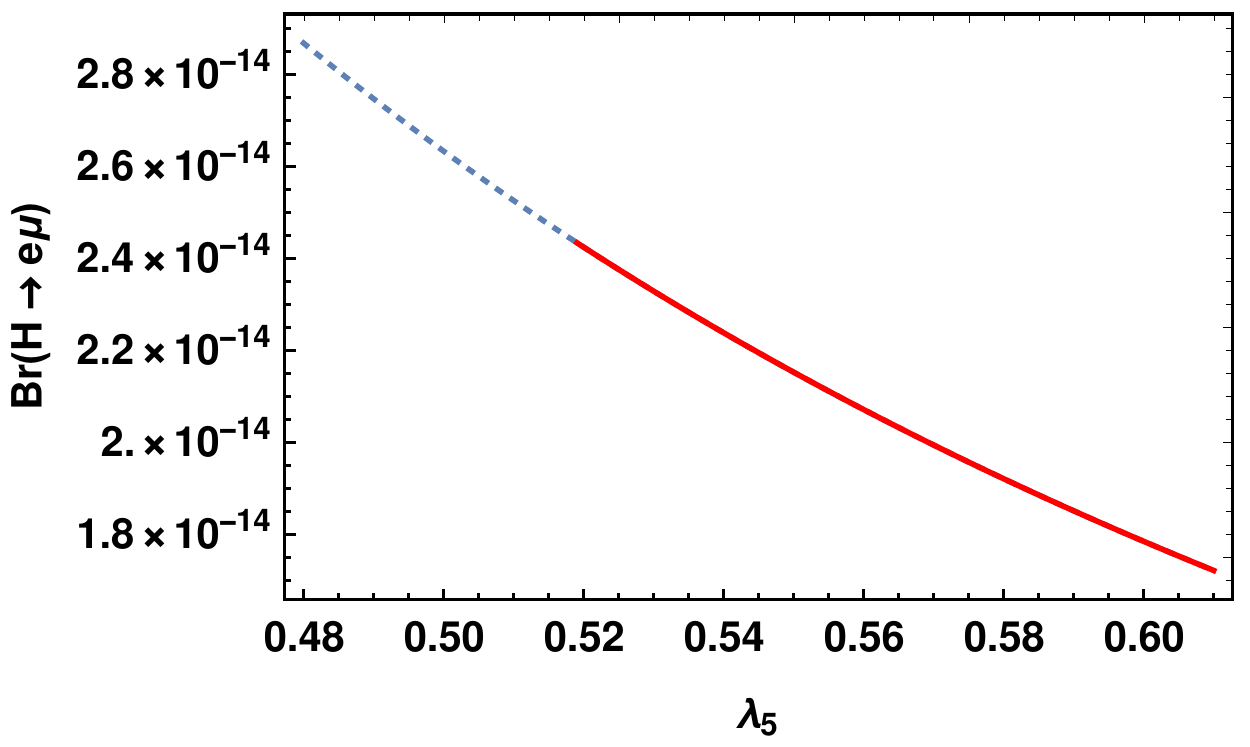}
\includegraphics[width=3.0in]{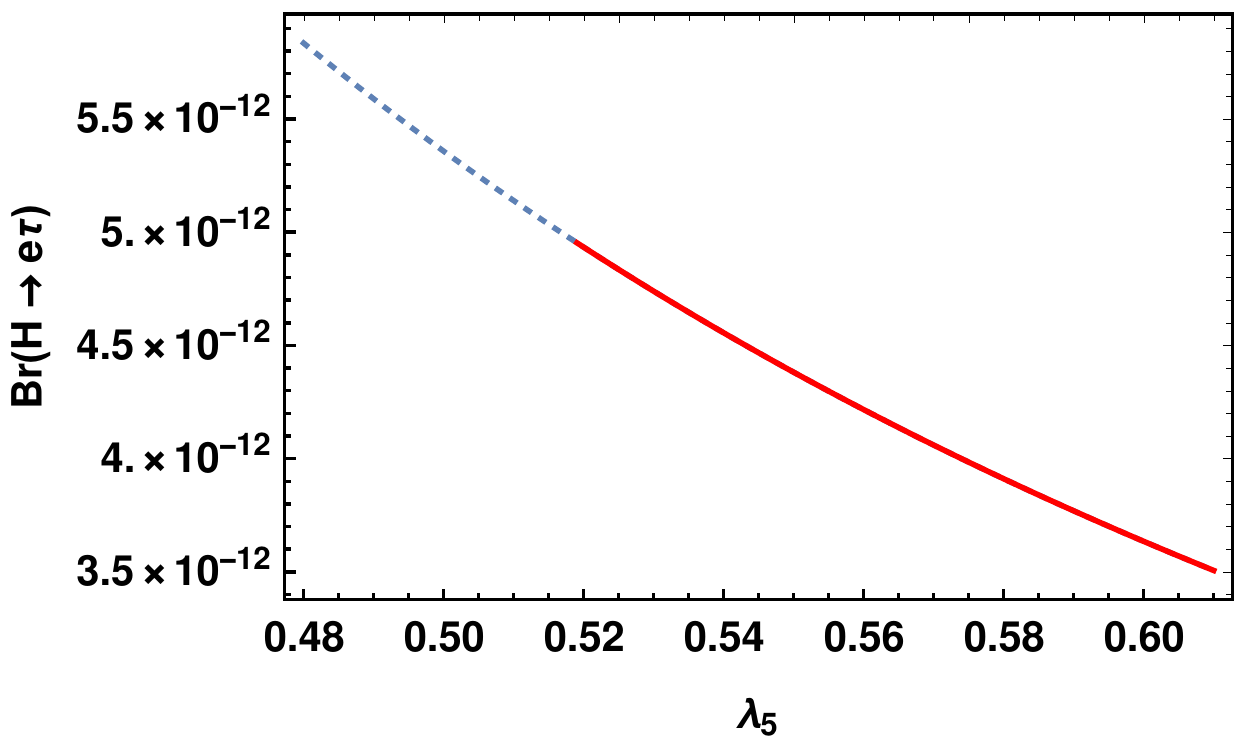}
\includegraphics[width=3.0in]{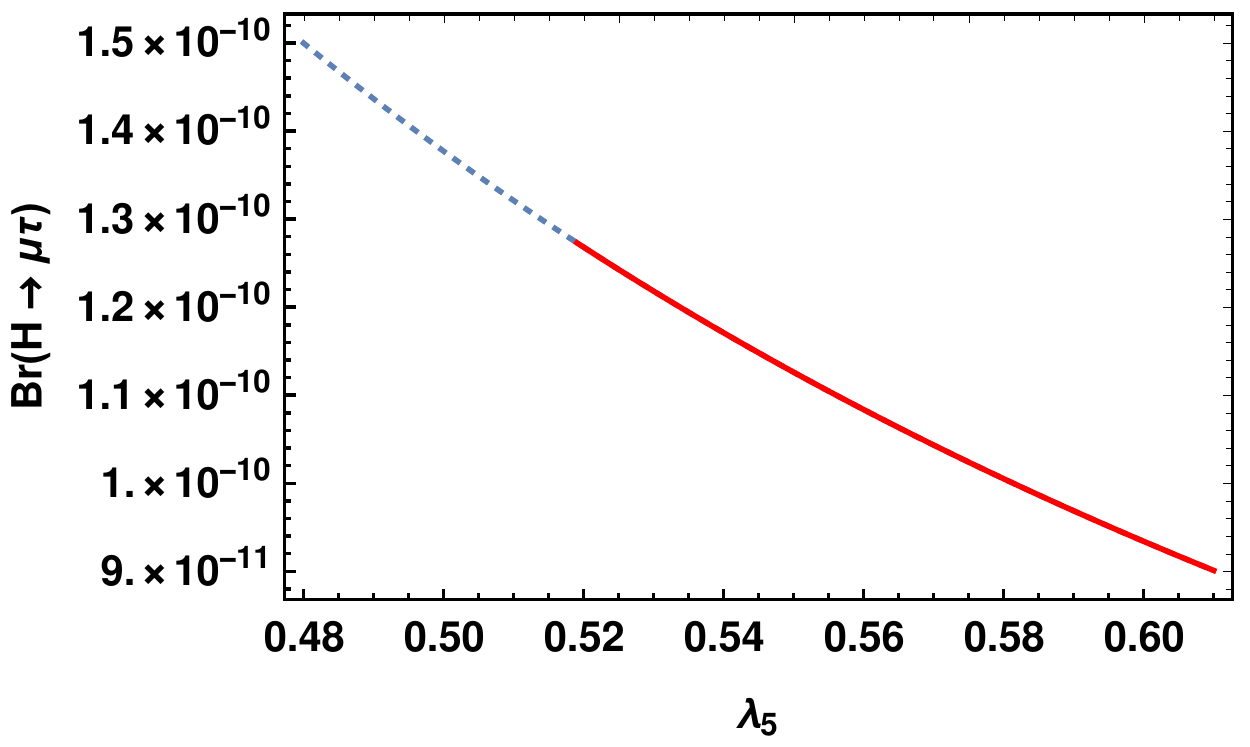}

\end{center}
\caption{Plots between ${\rm Br}(H\to\ell_\alpha\ell_\beta)$ and $\lambda_5$
for the case of IO, after applying the constraints from ${\rm Br}(\ell_\alpha
\to\ell_\beta\gamma)$. In these plots, solid lines are allowed and dotted lines
are excluded by the constraints due to
${\rm Br}(\ell_\alpha\to\ell_\beta\gamma)$.
See the caption of Fig. 5, for parametric values and
neutrino oscillation observables, which are used in these plots.}
\end{figure}
The masses for right-handed neutrinos are different in this figure as
compared to that in Fig. 7. Nevertheless, the allowed range of values for
${\rm Br}(H\to\ell_\alpha\ell_\beta)$ are found to be nearly same in Figs. 7
and 8.

Among the LFV decays of $Z$ and $H$, after applying the constraints from
${\rm Br}(\ell_\alpha\to\ell_\beta\gamma)$,
$H\to\mu\tau$ is found to have the largest branching
ratio, which is around $10^{-10}$. This indicates that probing LFV decays
of Higgs boson in experiments is one possible way to test the scotogenic
model. However, in our analysis of LFV decays of $H$, we have taken
$\lambda_3=4\pi$, which is the maximum possible value for this parameter.
In this model, the $\lambda_3$ coupling can also drive the decay
$H\to\gamma\gamma$. In the LHC experiment, it is found that there is no
enhancement in the signal strength of this decay as compared to the
standard model prediction \cite{pdg}. As a result of this, one can expect
some constraints on $\lambda_3$ parameter. Apart from this, the model
parameters of the scotogenic model can get additional constraints due
to precision electroweak observables and relic abundance of dark matter.
One may expect that the above mentioned constraints can lower the
allowed ranges for the branching ratios of LFV decays of $Z$ and $H$
in this model.

As stated in Sec. 1, in the context of scotogenic model, branchig ratio for
$H\to\mu\tau$ has been estimated as ${\rm Br}(H\to \mu\tau)\lapprox
10^{-7}\lambda_3^2$ \cite{hrs}, after applying the constraint from
${\rm Br}(\tau\to\mu\gamma)$. In our analysis, we have applied constraints
due to non-observation of all LFV decays of the form
$\ell_\alpha\to\ell_\beta\gamma$ and we have found that
${\rm Br}(H\to \mu\tau)$ can be as large as $\sim10^{-10}$, even with
$\lambda_3=4\pi$. Hence, our result
on ${\rm Br}(H\to \mu\tau)$ is more stringent than the above mentioned
estimation of Ref. \cite{hrs}.

\section{Conclusions}

In this work, we have studied LFV decays of $Z$ gauge boson and Higgs boson
in the scotogenic model. After deriving analytic expressions for the
branching ratios of the above mentioned decays, numerically we have studied
how large they can be in this model. The above mentioned numerical study
has been done by satisfying the following quantities:
fit to neutrino oscillation observables, constraints on
${\rm Br}(\ell_\alpha\to\ell_\beta\gamma)$ and perturbativity limits on the
parameters of the model. If we satisfy only the fit to
neutrino oscillation observables and the perturbativity limits on the model
parameters,
we have found the following maximum values for the branching ratios of
LFV decays of $Z$ and $H$:
${\rm Br}(Z\to e\mu,e\tau)\sim10^{-9}$, ${\rm Br}(Z\to \mu\tau)\sim10^{-8}$,
${\rm Br}(H\to e\mu)\sim10^{-7}$, ${\rm Br}(H\to e\tau)\sim10^{-4}$,
${\rm Br}(H\to \mu\tau)\sim10^{-3}$. However, in addition to satisfying
the above mentioned quantities, after satisfying constraints on
${\rm Br}(\ell_\alpha\to\ell_\beta\gamma)$, the above mentioned results
on the branching ratios get an additional suppression of about $10^{-7}$.
If the scotogenic model is true, results obtained in this work can
give indication about future results on LFV decays of $Z$ and $H$
in the upcoming experiments.

\vspace*{10mm}
\noindent
{\bf Note added}: While this manuscript was under preparation, Ref. \cite{note}
had appeared where LFV decays of Higgs boson were studied in the scotogenic
model. The method of computing the branching ratios for these decays and
numerical study done on them in Ref. \cite{note} are found to be different
from what we have done in this work.
After comparing ${\rm Br}(H\to\ell_\alpha\ell_\beta)$
versus ${\rm Br}(\ell_\alpha\to\ell_\beta\gamma)$ in Ref. \cite{note},
it is shown that the allowed values for
${\rm Br}(\ell_\alpha\to\ell_\beta\gamma)$ are suppressed to around $10^{-34}$.
Moreover, the branching ratio for $H\to\mu\tau$ is also shown
to be suppressed to around $10^{-37}$. The above mentioned results are
different from what we have presented here.

\end{document}